\documentclass[12pt]{article}
\usepackage{amsmath,amssymb,amsfonts}%
\usepackage{amsthm}%
\usepackage{graphicx}
\usepackage{subcaption}
\usepackage{enumerate}
\usepackage{algpseudocode}
\usepackage{algorithm}
\usepackage{rotating}
\usepackage{tikz}
\usepackage{makecell}
\usepackage{multirow}
\usepackage{booktabs}
\usepackage{xcolor}

\usepackage{natbib}
\bibpunct[, ]{(}{)}{;}{a}{}{,}
\renewcommand\bibfont{\fontsize{10}{12}\selectfont}

\theoremstyle{plain}
\newtheorem{theorem}{Theorem}
\newtheorem{lemma}{Lemma}
\newtheorem{proposition}{Proposition}
\newtheorem{corollary}{Corollary}[proposition]

\theoremstyle{definition}

\newtheorem{example}{Example}

\theoremstyle{remark}
\newtheorem{remark}{Remark}

\usepackage{url} 

\newif\ifblind
\blindfalse

\newif\iftrackchanges
\trackchangesfalse

\iftrackchanges
  \newenvironment{revised}
    {\color{blue}} 
    {}             
\else
  \newenvironment{revised}
    {} 
    {} 
\fi

\begin{document}




\ifblind
\begin{center}
    {\Large \bf Adaptive Bi-Level Variable Selection of Conditional Main Effects for Generalized Linear Models}\\
\end{center}
\else
    \begin{center}
        {\Large \bf Adaptive Bi-Level Variable Selection of Conditional Main Effects for Generalized Linear Models}\\
        \vskip 10pt
        {
        \vskip 5pt
        Kexin Xie and Xinwei Deng\footnote{Address for correspondence: Xinwei Deng, Professor, Department of Statistics,
            Virginia Tech, Blacksburg, VA 24061 (E-mail: xdeng@vt.edu).}\\
        \vskip 5pt
        Department of Statistics, Virginia Tech, USA
        }
    \end{center}
\fi

\begin{abstract}
Understanding interaction effects among variables is important for regression modeling in various applications. The conventional approach of quantifying interactions as the product of variables often lacks clear interpretability, especially in complex systems. The concept of conditional main effects (CME) provides a more intuitive and interpretable framework for capturing interaction effects by quantifying the effect of one variable conditional on the level of another. A recent method called {\it cmenet} further considered the bi-level selection of CMEs by leveraging their natural grouping structure (e.g., sibling and cousin groups) through penalization. However, there are several limitations in the cmenet method, including the coupling ability of penalties for within-group CMEs, lack of adaptiveness for between-group penalties, and restriction to linear models with continuous responses. To overcome these limitations, we propose an adaptive cmenet method for CME selection under the generalized linear model (GLM) framework. The proposed method considers a penalized likelihood approach with adaptive weights to enable effective bi-level variable selection, improving both between-group and within-group selection. An efficient algorithm for parameter estimation is also developed by employing an iteratively reweighted least squares procedure. The performance of the proposed method is evaluated by both simulation studies and real-data studies in gene association analysis. 
\end{abstract}

\noindent%
\vfill

\newpage
\section{Introduction}
\label{sec:intro}

In statistical analyses, the interactions between variables are of great interest for modeling and uncovering complex relationships among variables.
Traditionally, the interaction between two variables $x_{1}$ and $x_{2}$ is modeled as their product term $x_{1}x_{2}$ in regression models. However, such representations often lack intuitive interpretability, especially for non-specialists, making them challenging to apply in practice. The conditional main effects (CMEs), introduced by \cite{wu2015post}, offer a novel alternative by focusing on the effect of one variable while conditioning on the fixed level of another, i.e., $x_{1}|x_{2}$. The work of \cite{su2017cme} expanded on this idea by proposing a systematic strategy to identify an appropriate set of CMEs for the analysis of fractional factorial designs. The concept of CME offers an interpretable framework for understanding how certain factors may influence outcomes in the presence of specific conditions. For example, in gene association studies, identifying epistatic (gene-gene interaction) effects associated with multiple SNPs is crucial in these oligogenic/polygenic diseases \citep{shibata2024identification}. Instead of quantifying the interaction $AB$ between gene $A$ and gene $B$, the CME $A|B+$ provides a variable to quantify the significance of gene $A$ only when gene $B$ is present, thus enhancing biological interpretability.

However, when there are $p$ variables, even each with two levels, 
the total number of CMEs is $2p(p-1)$, which can be large even for a moderate $p$. Effective variable selection is therefore critical. 
Moreover, CMEs naturally exhibit structured groupings. Based on \cite{mak2019cmenet}, there are two general group structures \textit{sibling} groups and \textit{cousin} groups. 
Specifically, the sibling group of a ME \( A \) consists of CMEs sharing the same ME, such as \( \{A|B+, A|B-, A|C+, A|C-, \dots\} \). In contrast, the cousin group of \( A \) comprises CMEs that share the same conditional effect, including \( \{B|A+, B|A-, C|A+, C|A-, \dots\} \). 
Accommodating such group structures, it will be more appropriate to conduct the bi-level variable selection for identifying both active CME groups and the significant effects within these groups.
A pioneering work from \cite{mak2019cmenet} proposed a so-called {\it cmenet} method to consider the ``exponential-MC+" hierarchical penalization for bi-level variable selection in CMEs under the context of linear models.
Their method facilitates bi-level variable selection by controlling both \textit{between-group} and \textit{within-group} selection by adopting composition penalty in the penalized regression \citep{breheny2009penalized}.

Although the {\it cmenet} method has demonstrated good performance in selecting CMEs, several limitations remain. One issue is that the extent of coupling, where the penalty applied to a coefficient is reduced when grouped with other important predictors, is static and cannot adapt to the signal strength within the group, even with the inclusion of the coupling parameter~\(\tau\). 
Moreover, the tuning parameters \(\lambda_S\) (for sibling groups) and \(\lambda_C\) (for cousin groups) are not varying across all groups, limiting the flexibility in capturing group-specific variations. 
We also note that the {\it cmenet} method primarily works for linear models with continuous responses, making it unsuitable for situations where the response is binary, count-based, or follows a generalized linear model (GLM) structure.

To address these issues, we propose an \textit{adaptive cmenet} approach for bi-level CME selection under the GLM framework. 
Through systematic investigation of the penalty term used in the cmenet, we develop an adaptive penalization strategy to address the static coupling limitations in variable selection and extend the CME selection to generalized linear models. The key contributions of this work are as follows: First, inspired by adaptive Lasso \citep{zou2006adaptive}, the proposed method incorporates both group adaptive weights and individual adaptive weights to enable more accurate variable selection of CMEs. 
The group adaptive weights mitigate the static coupling issue by assigning higher weights to effects within the same groups that are likely to be coupled. 
The individual adaptive weights can regulate the shrinkage rate, enabling more accurate coefficient estimation. A theoretical interpretation of the adaptive weights is provided to justify the choice of weights. 
Furthermore, we extend the CME selection to GLMs and develop an efficient iteratively reweighted least squares (IRLS) algorithm \citep{green1984iteratively} for parameter estimation. 
The proposed IRLS algorithm incorporates the coordinate descent methods \citep{fu1998penalized, friedman2007pathwise, mazumder2011sparsenet} and a majorization-minimization technique \citep{hunter2000quantile} to have a closed-form solution for the optimization in each iteration. 
The implementation of the proposed method is deployed through the R package, \texttt{glmcmenet}, and will be publicly available for use in practical applications.

The rest of this article is organized as follows. We describe the concept of the CMEs briefly and review the cmenet method in Section \ref{sec:cme}.  
We detail the proposed adaptive cmenet approach for GLMs in Section \ref{sec:adap} and present the parameter estimation algorithm in Section \ref{opt_algt}. 
The performance of the adaptive cmenet is illustrated in Section \ref{sec:sim} via several numerical simulations.  The application to the gene-association studies is analyzed in Section \ref{sec:gene}. Section \ref{sec:conc} concludes this work with some discussions on future works.

\section{Background and Motivation}

\subsection{Review of Conditional Main Effects}
\label{sec:cme}

The definition of conditional main effects (CME) is first introduced by \citet{wu2015post}. Following the work of \citet{mak2019cmenet}, we use the definition of CME derived from main effects (ME). Suppose that there are \(n\) observations with \(p\) factors of 2-level under consideration, represented by binary covariate vector \( \mathbf{x}_j=(x_{1,j},x_{2,j},\dots,x_{n,j})^T\in\{-1,1\}^n \), \(j=1,\dots,p\) in the model. Briefly, for the covariate \(x_{i,j}\) conditional on \(x_{i,k}\) at level \(+1\) for object \(i\), the CME covariate is defined as \(x_{i,j|k+} = x_{i,j}\) when \(x_{i,k} = +1\), and \(x_{i,j|k+} = 0\) when \(x_{i,k} = -1\). The CME covariate for \(x_{i,j|k-}\) is defined analogously (see \citet{mak2019cmenet} for further details). For convenience, we adopt the notation \(A, B, C, D, \dots\) to represent the MEs that quantify the effect of covariates \(\mathbf{x}_1, \mathbf{x}_2, \dots, \mathbf{x}_p\), respectively. Similarly, the CMEs that quantify the effect of covariates \(\mathbf{x}_{j|k+}\) and \(\mathbf{x}_{j|k-}\) are denoted as \(J|K+\) and \(J|K-\).  In the CMEs \(J|K+\) and \(J|K-\), the effects $J$ and $K$ are referred as the \textit{parent effect} and the \textit{child effect}, respectively. This notation will be used consistently throughout the remainder of the paper.

Note that the conventional interaction terms are often not straightforward for interpretation.
Consider a simple example involving two covariates, \(A\) and \(B\). 
Traditionally, their interaction is represented as the product term \(AB\).
However, this term \(AB\) alone does not provide explicit information about the individual values or levels of \(A\) and \(B\), making it difficult to disentangle their interaction in practical terms. For instance, \(AB = 1\) could result from either \(A = +1, B = +1\) or \(A = -1, B = -1\). CMEs address this ambiguity by capturing conditional relationships more explicitly. Table \ref{tab:CME_AB} presents the four possible CMEs derived from these MEs.
\begin{table}[t]
\caption{Model Matrix for the two MEs A and B, with its interaction effect and four CMEs.}\label{tab:CME_AB}
\begin{center}
    \begin{tabular}{rr|r|rrrr}
    \hline
    $A$ & $B$ & $AB$ & $A|B+$ & $A|B-$ & $B|A+$ & $B|A-$ \\ 
    \hline
    +1 & +1 & +1 & +1 & 0 & +1 & 0\\
    +1 & -1 & -1 & 0 & +1 & -1 & 0\\
    -1 & +1 & -1 & -1 & 0 & 0 & +1\\
    -1 & -1 & +1 & 0 & -1 & 0 & -1\\
    \hline
    \end{tabular}                    
\end{center}
\end{table}

It is evident that there are complex hierarchical structures among the ME covariates and CME covariates, either as parent terms or as child terms.
Following the grouping structures in \citet{mak2019cmenet}, there are three general categories: (1) \textit{Sibling} CMEs, which share the same parent effects, such as $\{A|B+, \allowbreak A|B-, \allowbreak A|C+, \allowbreak A|C-, \cdots\}$; (2) \textit{Cousin} CMEs, which share the same child effects, such as $\{B|A+,\allowbreak B|A-,\allowbreak C|A+,\allowbreak C|A-, \cdots\}$; and (3) \textit{Parent-Child} pairs, such as $\{A, A|B+\}$, $\{A, A|B-\},\ \dots$. \citet{mak2019cmenet} examined the collinearity and pairwise correlations within these groups and observed that the correlation magnitudes create a natural hierarchy among effect groups, leading to inconsistent selection when using standard variable selection methods. To address this, they proposed a bi-level variable selection method for CMEs, implemented in the cmenet framework. This idea builds upon earlier developments in bi-level variable selection, such as the group bridge penalty by \citet{huang2009group}, which applies a bridge penalty to the \(\ell_1\)-norm of predefined groups to encourage group-level sparsity. \cite{breheny2009penalized} later introduced a general framework for composite penalties, which employs an \textit{outer} penalty to regulate a sum of \textit{inner} penalties. It facilitates bi-level variable selection by controlling both between-group and within-group selection. 

Let \((\mathbf{y},\mathbf{X})\) represent the complete set of observations, where \(\mathbf{y} \in \mathbb{R}^{n \times 1}\) is the response variable corresponding to the full model matrix \(\mathbf{X} = (\mathbf{x}_1, \mathbf{x}_2, \dots, \mathbf{x}_p, \dots, \mathbf{x}_{p'}) \in \mathbb{R}^{n \times p'}\). Here, \(\{\mathbf{x}_j\}_{j=1}^p\) and \(\{\mathbf{x}_j\}_{j=p+1}^{p'}\) are the normalized vectors for the \(p\) MEs and the \(4{p \choose 2}\) CMEs, respectively. Define \(\mathcal{S}(j) = \{J, J|A+, J|A-, J|B+, J|B-, \cdots\}\) as the sibling group for the parent effect \(j\), and \(\mathcal{C}(j) = \{J, A|J+, A|J-, B|J+, B|J-, \cdots\}\) as the cousin group for the child effect \(j\), \(j = 1, \dots, p\). The parameter estimation is then conducted as follows:
\begin{equation}
\begin{aligned}
 \underset{\boldsymbol{\beta}}{\min}\ Q(\boldsymbol{\beta})&=\underset{\boldsymbol{\beta}}{\min} \left\{ \frac{1}{2}\left \|\boldsymbol{y-X\beta}\right \|_2^2+P_S(\boldsymbol{\beta})+P_C(\boldsymbol{\beta}) \right\}, \\
 P_\mathcal{S}(\boldsymbol{\beta})=\sum_{j=1}^p f_{out,\mathcal{S}}&\left\{ \underset{k\in \mathcal{S}(j)}{\sum}f_{in,\mathcal{S}}(\beta_k) \right\},
 P_\mathcal{C}(\boldsymbol{\beta})=\sum_{j=1}^p f_{out,\mathcal{C}}\left\{ \underset{k\in \mathcal{C}(j)}{\sum}f_{in,\mathcal{C}}(\beta_k) \right\},
\end{aligned}\label{equ1}
\end{equation}
where \(f_{out,\mathcal{S}}\) and \(f_{in,\mathcal{S}}\) (\(f_{out,\mathcal{C}}\) and \(f_{in,\mathcal{C}}\)) represent the outer and inner penalties, which govern the between-group and within-group selection for sibling (or cousin) groups, respectively. Using different combinations of outer and inner penalties, 
\citet{mak2019cmenet} adopted the exponential penalty \(\eta_{\lambda_s,\tau}(\theta)\) (or \(\eta_{\lambda_c,\tau}(\theta)\)) for the outer penalty \citep{breheny2015group} and the scaled minimax concave-plus (MC+) penalty \(g_{\lambda_s,\gamma}(\theta)\) (or \(g_{\lambda_c,\gamma}(\theta)\)) for the inner penalty \citep{zhang2010nearly}. These penalties are defined as follows:
\begin{equation}
\begin{aligned}
\eta_{\lambda,\tau}(\theta) &= \frac{\lambda^2}{\tau}\left\{ 1-\exp(-\frac{\tau\theta}{\lambda})\right\},\\
\mathrm{and} \quad g_{\lambda,\gamma}(\beta) &= \int_0^{|\beta|}\left(1-\frac{x}{\lambda\gamma}\right)_+ dx,
\end{aligned}\label{equ2}
\end{equation}
where \(\lambda\), \(\gamma\), and \(\tau\) are tuning parameters that regulate the degree of regularization,  nonconvexity, and coupling, respectively. Specifically, two penalization parameters \(\lambda_s>0\) and \(\lambda_c>0\) are the penalization parameters applied to sibling and cousin groups, remaining constant across all groups. The MC+ penalty incorporates the parameter \(\gamma > 1\), which acts as a transition between the \(\ell_0\)-norm penalty (approximated as \(\gamma \to 1\)) and the \(\ell_1\)-norm penalty in LASSO (approximated as \(\gamma \to \infty\)). The coupling parameter \(\tau\) within the exponential penalty governs the strength of the relationship among variables within the same group. These penalty functions can effectively control group-wise coupling and sparsity, addressing challenges in bi-level variable selection for CMEs.

\begin{revised}
\subsection{Real-World Motivation}
Biologists often observe epistasis, where the effect of a gene (or genetic variant) on a phenotype depends on the presence of another variant or on whether a pathway is active. A main-effects screen can miss such patterns, while a full search over all interactions can be very large and hard to interpret. Conditional main effects (CMEs) can address this gap. A statement such as ``SNP $J$ given SNP $K=1$" expresses the biological context directly and groups related effects in a way that scientists use. In the biological context, the SNP whose effect is being evaluated ($J$) is the focal locus, and the SNP on which it is conditioned ($K$) is the modifier locus. This representation clarifies when and under what genetic context a locus influences the phenotype, while keeping the model concise and interpretable. To illustrate how CMEs capture such context-dependent effects in practice, we consider a genome-wide association study in Arabidopsis thaliana.

We study the visual chlorosis phenotype from the GWAS of \citet{atwell2010genome} (details in Section~\ref{sec:gene}). Arabidopsis is a predominantly self-fertilizing model plant with substantial genetic variation across many traits. CMEs can help connect SNPs to specific biological pathways. For example, a CME might show that a SNP near a chlorophyll gene only has an effect if a variant in an iron-related gene is also present. Another CME might show that a redox-enzyme SNP has a stronger effect when its antioxidant partner gene is absent. This type of ``conditional" statement facilitates comparison with existing functional annotations, and is associated with experimentally testable hypotheses. Meanwhile, the \textit{cmenet} method also shows its limitations in this data scenario. First, the chlorosis phenotype is a binary outcome, and the \textit{cmenet} method, built on a linear Gaussian model, cannot model binary responses on an appropriate scale.  
Second, genetic markers are often correlated. When \textit{cmenet} finds one strong CME, it often selects a group of correlated, nearby CMEs as well. This creates redundancy in the results, populating the output with long lists of CMEs that all point to the same underlying genetic source and reducing interpretability.

Our framework is designed to address these challenges. We keep the CME representation and consider a logistic model for the analysis of a binary trait. We then adopt the penalization at both the group level and the individual level. The key advantage is that the relative influence of the two CME groups shifts with the evidence rather than being fixed in advance. It can obtain an interpretable set of context-specific associations that can highlight candidate gene–gene interaction patterns for follow-up study. The details of the proposed method will be introduced in the following Section~\ref{sec:problem}.
\end{revised}

\section{The Proposed Method: Adaptive cmenet}
\label{sec:adap}

In this section, we first investigate the coupling issue in the cmenet method and identify its limitations in Section \ref{sec:problem}. 
We then propose a new penalization approach in Section \ref{adp_cmenet} to address these limitations. Finally, we examine the effectiveness of the proposed approach in comparison with the cmenet method in Section \ref{sec:coupling_adaptive}.

\subsection{Coupling Ability in the cmenet}
\label{sec:problem}

Note that the choice of the ``exponential-MC+" penalty framework is motivated by two key principles: (1) \textit{CME coupling}, which states that a CME is more likely to be included in the model if other effects within the same sibling or cousin group have already been selected; and (2) \textit{CME reduction}, which implies that a ME is more easily selected when effects in its sibling or cousin groups have already entered the model. To elucidate these principles, we adopt the concept of the ``selection threshold" introduced by \cite{breheny2015group}, which refers to the minimum level of association between a covariate and the outcome that is necessary for the covariate to be selected in the model. Let \(z\) represent the unpenalized estimate for \(\beta_{j}\), i.e, \(z=\frac{1}{n}\mathbf{x}_{j}^T(\mathbf{y}-\mathbf{X}_{-j}\mathbf{\beta}_{-j})\)). Here, \( \mathbf{X}_{-j} \in \mathbb{R}^{n \times (p'-1)} \) represents the design matrix excluding the \( j \)th covariate, and \( \boldsymbol{\beta}_{-j} \in \mathbb{R}^{(p'-1)} \) is the corresponding coefficient vector. The estimate \( z \) corresponds to the value that minimizes the loss function while keeping all other coefficients fixed. The ``selection threshold" is formally defined as \(t_{\beta_{j}} := \inf\{|z|: \hat{\beta}_{j} \neq 0\}\) for \(j=1,\dots,p'\). In other words, it represents the infimum of all values of $|z|$ for which the corresponding coefficient $\hat{\beta}_{j}$ is nonzero. Derived from \eqref{equ1}, the selection threshold \( t_{\beta_{j|k+}}\) for $\beta_{j|k+}$ is expressed as:
\begin{equation}
t_{\beta_{j|k+}} = \lambda_s\exp\{-\frac{\tau \Vert \boldsymbol{\beta}_{\mathcal{S}(j)} \Vert_{\lambda_s,\gamma}}{\lambda_s}\} + \lambda_c\exp\{-\frac{\tau \Vert \boldsymbol{\beta}_{\mathcal{C}(k)} \Vert_{\lambda_c,\gamma}}{\lambda_c}\}.
\label{equ3}
\end{equation}
Here, the coefficient vector of a group \(\mathcal{G}\) is denoted by \(\boldsymbol{\beta}_{\mathcal{G}} \in \mathbb{R}^{|\mathcal{G}|}\), and the penalty term is defined as \(\Vert \boldsymbol{\beta}_{\mathcal{G}}\Vert_{\lambda_\mathcal{G}, \gamma} := \sum_{l\in\mathcal{G}}g_{\lambda_{\mathcal{G}},\gamma}(\beta_l)\). The selection threshold \(t_{\beta_j}\) for the coefficient \(\beta_{j}\) is defined in an analogous manner. For simplicity, we introduce the notations \(\Delta_{\mathcal{S}(j)} := \lambda_s\exp\{-\frac{\tau \Vert \boldsymbol{\beta}_{\mathcal{S}(j)} \Vert_{\lambda_s,\gamma}}{\lambda_s}\}\) and \(\Delta_{\mathcal{C}(k)} := \lambda_c\exp\{-\frac{\tau \Vert \boldsymbol{\beta}_{\mathcal{C}(k)} \Vert_{\lambda_c,\gamma}}{\lambda_c}\}\), which will be used throughout the remainder of the paper.

\begin{revised}
Let us consider a model in which \(A|B+\) has been selected. Two candidate effects are under consideration: \(A|C+\) and \(A\). Figure \ref{fig:1}(a) illustrates how changes in \(\beta_{A|B+}\) will influence the selection thresholds for \(\beta_{A|C+}\) and \(\beta_{A}\), respectively. When no variables are initially included in the model, the selection threshold remains high regardless of the coefficient magnitude. However, as a sibling variable (\(\mathbf{x}_{A|\cdot}\)) or a cousin variable (\(\mathbf{x}_{\cdot|C\cdot}\) for \(A|C+\), or \(\mathbf{x}_{\cdot|A\cdot}\) for \(A\)) is included into the model, the selection thresholds for both \(\beta_{A|C+}\) and \(\beta_{A}\) decrease. This reduction is particularly pronounced when both sibling and cousin effects are included, facilitating the inclusion of additional effects within the same group. This phenomenon underpins the first principle: \textit{CME Coupling}. To contrast with the behavior of the sibling effect, Figure \ref{fig:1}(b) considers a model that includes \(B|A+\) and varies its coefficient \(\beta_{B|A+}\). The selection threshold for \(\beta_{A}\) (represented by dotted lines) declines with \(\beta_{B|A+}\), whereas the threshold for \(\beta_{A|C+}\) (represented by solid lines) remains essentially unchanged because cousins of $C$ are fixed. The lower threshold value for the \(\beta_{A}\) indicates that the ME $A$ is more likely to be included in the model than the CME \(A|C+\). That is, when effects involving $A$ have already been selected from $\mathcal{S}(A)$ or from $\mathcal{C}(A)$, the ME $A$ enters the model more easily. When many such effects are selected, the set of conditional effects tends to contract to the underlying main effect. This behavior exemplifies the second principle: \textit{CME Reduction}.

\begin{figure}[t]
\begin{center}
\includegraphics[width=1\textwidth]{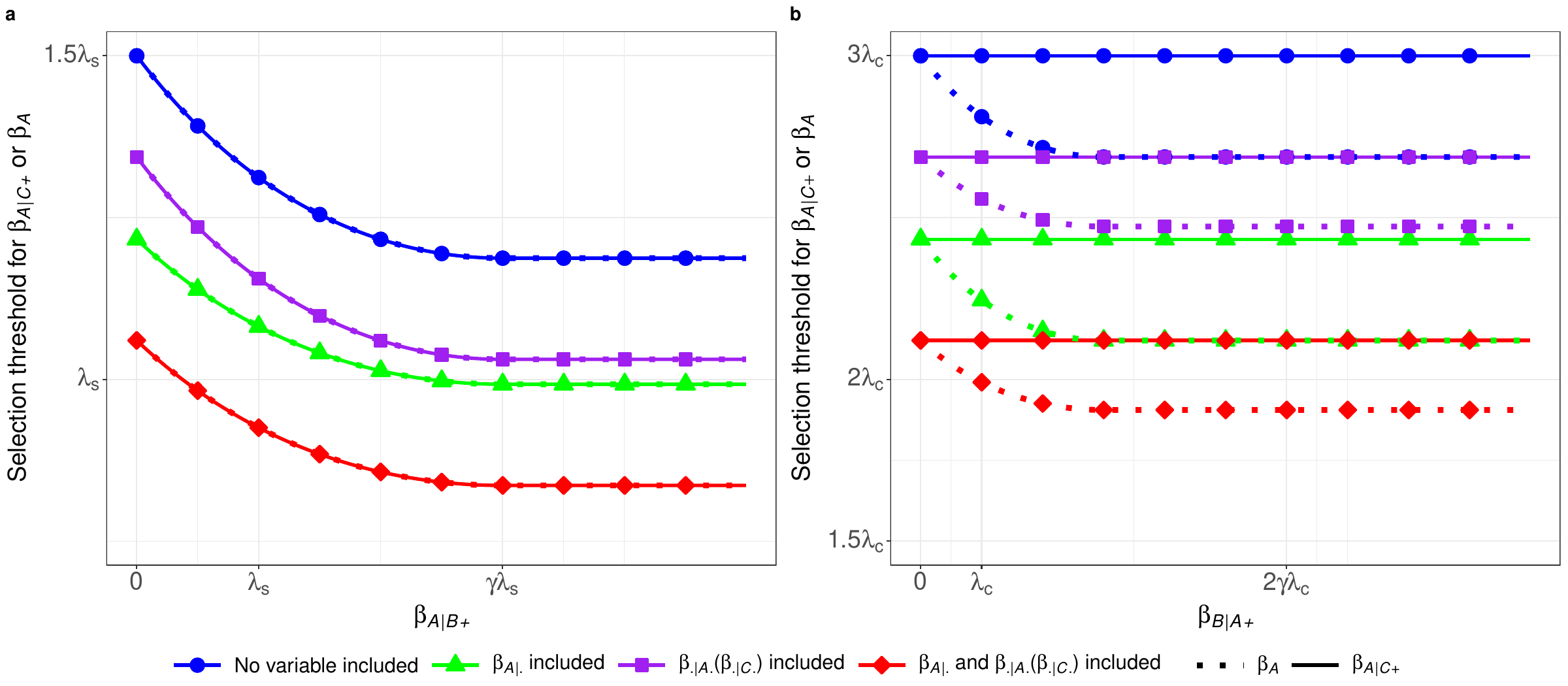}
\end{center}
\caption{\begin{revised} 
Effect of varying CME coefficients on selection thresholds of \textit{cmenet}: (a) Vary $\beta_{A|B+}$; (b) Vary $\beta_{B|A+}$. Curves show the threshold for $\beta_{A|C+}$ (dotted line) and $\beta_{A}$ (solid line) under four contexts: when no other variable included (blue), one sibling included (green), one cousin included (purple), and one sibling and one cousin included (red) in the model. Baseline \texttt{cmenet} setting: $(\lambda_s,\lambda_c,\gamma,\tau)=(1, 0.5, 3, 0.25)$. The adding variable has the same coefficient as $\beta=2$. \label{fig:1}
\end{revised}}
\end{figure}

An important observation is that as the coefficient magnitude increases, the selection threshold stabilizes at a fixed level. That is, the extent of coupling is inherently controlled by group size and the coupling parameter \(\tau\). This phenomenon stems from the characteristics of the inner penalty function \(g_{\lambda,\gamma}(\beta)\) in \eqref{equ2}, which is monotonic increasing and gradually reduces penalization as \(|\beta|\) increases, up to the critical point \(|\beta| = \gamma\lambda\). Beyond this critical value, \(g_{\lambda,\gamma}(\beta)\) becomes constant at its maximum value of \(\gamma\lambda/2\). As a result, the selection threshold reaches a plateau at a fixed value of \(\eta'_{\lambda,\tau}(\theta)|_{\theta = \gamma\lambda/2}\), remaining unaffected by any further increases in the coefficient magnitude. This plateau prevents excessive coupling because each term in the exponent in $\eta_{\lambda,\tau}(\theta)$ has a finite contribution; once terms have saturated, enlarging coefficients cannot push the threshold downward without bound.

Although this plateau is desirable at very large signals, the method suffers from a static coupling limitation, which manifests in two practical ways. 
First, it has a fixed between-group relative influence determined by the non-adaptive parameters $\lambda_s$ and $\lambda_c$. This balance is locked in from the start and cannot adapt to the data. For example, in Figure~\ref{fig:1}, setting $\lambda_s=2\lambda_c$ causes the green curve to lie below the purple one, regardless of how much signal is present. This fixed setting locks in a predefined ratio between the groups, making the threshold insensitive to changes in the actual strength of evidence. 
Second, and more importantly, it has a constant marginal effect of adding a new variable within a group, regardless of the current strength of that group. It is clearly shown in Figure \ref{fig:1} (a): the gap between threshold curves, such as between the blue and green lines (or between purple and red), remains unchanged as $\beta_{A|B+}$ increases. The lines run parallel. Moreover, due to the uniform bounds of the inner penalty, the benefit of adding a second sibling is exactly the same as adding the first. Consequently, the threshold keeps dropping as you add more siblings, even if the group is already ``saturated". This lack of diminishing marginal returns can lead to overemphasis on already dominant groups and insufficient responsiveness to emerging signals in competing groups. 

These considerations suggest that it is important to allow the penalization to be adaptive to the current level of signal of each group. First, to address the fixed between-group relative influence, it will be more desirable to let the main penalization parameters ($\lambda_s$ and $\lambda_c$) adjust based on the total signal strength of their respective groups. Thus, the procedure can dynamically ``rebalance" the influence of the sibling and cousin groups. 
Second, to address the constant within-group marginal effect, we need to adjust the inner penalty for each coefficient adaptively. This means the inner penalty applied to a coefficient (e.g., $\beta_{A|C+}$) should decrease as the signal from a related coefficient ($\beta_{A|B+}$) increases. This makes the selection threshold drop earlier in response to initial weak signals within the group, and mitigates the marginal impact once the group already contains large signals.
Thus, we develop an adaptive penalization framework, which is described in detail in the following section.
\end{revised}

\subsection{The Proposed Adaptive cmenet and \texttt{glmcmenet}}
\label{adp_cmenet}

As discussed in Section \ref{sec:problem}, the issue of static coupling limitation primarily arises from the uniform bounds of the inner penalty and the non-adaptive penalization parameters applied across groups and individuals. Drawing inspiration from the adaptive Lasso in \citet{zou2006adaptive} and the adaptive sparse group Lasso developed by \citet{fang2015bi}, we consider incorporating adaptive weights for different groups and individual effects. Accordingly, the proposed adaptive cmenet penalization is defined as follows:
\begin{equation}
\tilde{P}_\mathcal{G}(\boldsymbol{\beta})=\sum_{j=1}^p f^{(j)}_{out,\mathcal{G}}\Biggl\{ \underset{k\in \mathcal{G}(j)}{\sum}\omega_kf^{(j)}_{in,\mathcal{G}}(\beta_k^{(j)}) \Biggr\},\quad \mathcal{G}\in\{\mathcal{S},\mathcal{C}\} \ \text{and} \ k \in \mathcal{G}(j)
\label{equ4}
\end{equation}
where $f^{(j)}_{out,\mathcal{G}}(\theta)=\eta_{\lambda_{\mathcal{G}(j)},\tau}(\theta)$ and $f^{(j)}_{in,\mathcal{G}}(\beta)=g_{\lambda_{\mathcal{G}(j)},\gamma}(\beta)$. Here $\lambda_{\mathcal{G}(j)}$ is the adaptive weight for group $\mathcal{G}(j)$, and $\omega_k$ is the individual weight for ME or CME within group $\mathcal{G}(j)$. The rationale of introducing adaptive weights is: $\lambda_{\mathcal{G}(j)}$ resolves the issue of non-adaptive penalization parameters across groups, while $\omega_k$ relaxes the uniform boundary of the inner penalty. To illustrate the distinction between the cmenet penalty and the proposed adaptive cmenet penalty, readers can consider a simplified example as illustrated in Example \ref{examp}. Consequently, the parameter estimation can be expressed as:
\begin{equation}
\underset{\boldsymbol{\beta}}{\min}\ Q(\boldsymbol{\beta})=\underset{\boldsymbol{\beta}}{\min} \Bigl\{ \frac{1}{2}\left \|\mathbf{y-X}\boldsymbol{\beta}\right \|_2^2+\tilde{P}_S(\boldsymbol{\beta})+\tilde{P}_C(\boldsymbol{\beta}) \Bigr\}.
\label{equ5}
\end{equation}

\begin{example}
Consider a simple model with only two MEs \(A\) and \(B\), then the corresponding sibling groups are \(\mathcal{S}(A)=\{A,A|B+,A|B-\}\), \(\mathcal{S}(B)=\{B,B|A-,B|A+\}\). For the cmenet penalty in \eqref{equ1}, the penalization for sibling groups is defined as:
\begin{equation*}
\begin{aligned}
P_\mathcal{S}(\boldsymbol{\beta})=\frac{\lambda^2}{\tau}\Biggl\{ 1-\exp\biggl[-\frac{\tau}{\lambda}\Bigl\{g_{\lambda,\gamma}(\beta_{A})+g_{\lambda,\gamma}(\beta_{A|B+})+g_{\lambda,\gamma}(\beta_{A|B-})\Bigr\}\biggr]\Biggr\}+\\
\frac{\lambda^2}{\tau}\Biggl\{ 1-\exp\biggl[-\frac{\tau}{\lambda}\Bigl\{g_{\lambda,\gamma}(\beta_{B})+g_{\lambda,\gamma}(\beta_{B|A+})+g_{\lambda,\gamma}(\beta_{B|A-})\Bigl\}\biggl]\Biggr\}.
\label{equ_expcme}
\end{aligned}
\end{equation*}
In contrast, for the proposed adaptive cmenet penalty in \eqref{equ4}, sibling group  weights $\mathcal{S}(j)$ and individual weights $\omega_k$ are introduced:
\begin{equation*}
\begin{aligned}
\begin{split}
\tilde{P}_\mathcal{S}(\boldsymbol{\beta})= &\frac{\lambda_{\mathcal{S}(A)}^2}{\tau}\Biggl\{ 1-\exp\biggl[-\tau\Bigl\{\frac{\omega_{A}}{\lambda_{\mathcal{S}(A)}}\cdot g_{\lambda_{\mathcal{S}(A)},\gamma}(\beta_{A})+\\ 
& \qquad \qquad \frac{\omega_{A|B+}}{\lambda_{\mathcal{S}(A)}}\cdot g_{\lambda_{\mathcal{S}(A)},\gamma}(\beta_{A|B+})+\frac{\omega_{A|B-}}{\lambda_{\mathcal{S}(A)}}\cdot g_{\lambda_{\mathcal{S}(A)},\gamma}(\beta_{A|B-})\Bigr\}\biggr]\Biggr\}+ \\
& \frac{\lambda_{\mathcal{S}(B)}^2}{\tau}\Biggl\{ 1-\exp\biggl[-\tau\Bigl\{\frac{\omega_{B}}{\lambda_{\mathcal{S}(B)}}\cdot g_{\lambda_{\mathcal{S}(B)},\gamma}(\beta_{B})+ \\
& \qquad \qquad \frac{\omega_{B|A+}}{\lambda_{\mathcal{S}(B)}}\cdot g_{\lambda_{\mathcal{S}(B)},\gamma}(\beta_{B|A+})+\frac{\omega_{B|A-}}{\lambda_{\mathcal{S}(B)}}\cdot g_{\lambda_{\mathcal{S}(B)},\gamma}(\beta_{B|A-})\Bigr\}\biggr]\Biggr\}.
\label{equ_expadpcme}
\end{split}
\end{aligned}
\end{equation*}
A similar formula can be written for cousin group penalties. 
\label{examp}
\end{example}
For each group \(\mathcal{G}(j)\), we consider the group weight to be \(\Omega_{\mathcal{G}(j)}=(\|\boldsymbol{\hat{\beta}}^{ridge}_{\mathcal{G}(j)}\|_1 + \frac{1}{n})^{-1}\), where \(\|\boldsymbol{\hat{\beta}}^{ridge}_{\mathcal{G}(j)}\|_1\) denotes the \(l_1\)-norm of the ridge estimators within group \(\mathcal{G}(j)\). Thus, the adaptive penalization parameter of this group is \(\lambda_{\mathcal{G}(j)} = \lambda_{\mathcal{G}}\)\(\Omega_{\mathcal{G}(j)}\). The individual weight within this group is defined as \(\omega_k = (|\hat{\beta}^{ridge}_k| + \frac{1}{n})^{-1}, k \in \mathcal{G}(j)\). The use of ridge estimators for weight construction aligns with \citet{zou2006adaptive}. Given the presence of implicit multicollinearity in CME group structures, ridge estimation is preferred over \(\hat{\beta}^{MLE}\) due to its superior consistency and robustness. To prevent division by zero, we follow \citet{fang2015bi} and add \(1/n\) to the denominator. The adaptive cmenet method preserves the principles of CME coupling and CME reduction. By incorporating adaptive weights, this approach effectively alleviates the issue of static coupling, as further elaborated in Section~\ref{opt_algt}.

Note that the cmenet method is primarily designed for linear models.
To broaden the scope of CMEs, we extend the adaptive cmenet framework to generalized linear models (GLM), hereafter named \texttt{glmcmenet}.
For GLM, the loss function is typically the negative log-likelihood derived from the exponential family. GLMs extend linear models to accommodate various types of response variables, by linking the mean response to the predictors through a link function. Let \(\eta_i = \mathbf{x}_i^T \boldsymbol{\beta}\) denote the linear predictor, where \(\mathbf{x}_i\) is the vector of predictors for the \(i\)th observation. The relationship between the response mean \(\mu_i\) and the linear predictor \(\eta_i\) is specified by a canonical link function \(g(\mu_i) = \eta_i\). The loss function (i.e., the negative log-likelihood function) for GLMs is $-\frac{1}{n} \sum_{i=1}^n \left[y_i \cdot \eta_i - b(\eta_i) + c(y_i)\right]$,
where \(b(\eta_i)\) and \(c(y_i)\) are functions defined by the exponential family distribution of the response variable. 
The optimization for the proposed method can be written as 
\begin{equation}
\underset{\boldsymbol{\beta}}{\min}\ Q(\boldsymbol{\beta})=\underset{\boldsymbol{\beta}}{\min} \Bigl\{ -\frac{1}{n} \sum_{i=1}^n \left[y_i \cdot \eta_i - b(\eta_i) + c(y_i)\right]+\tilde{P}_S(\boldsymbol{\beta})+\tilde{P}_C(\boldsymbol{\beta}) \Bigr\},
\label{acme-glmnet}
\end{equation}
where \(\tilde{P}_S(\boldsymbol{\beta})\) and \(\tilde{P}_c(\boldsymbol{\beta})\) are defined in \eqref{equ1}. The details on parameter estimation will be described in Section \ref{opt_algt}.

\subsection{Coupling Ability of the Adaptive cmenet}
\label{sec:coupling_adaptive}

\begin{figure}[t]
\begin{center}
\includegraphics[width=1\textwidth]{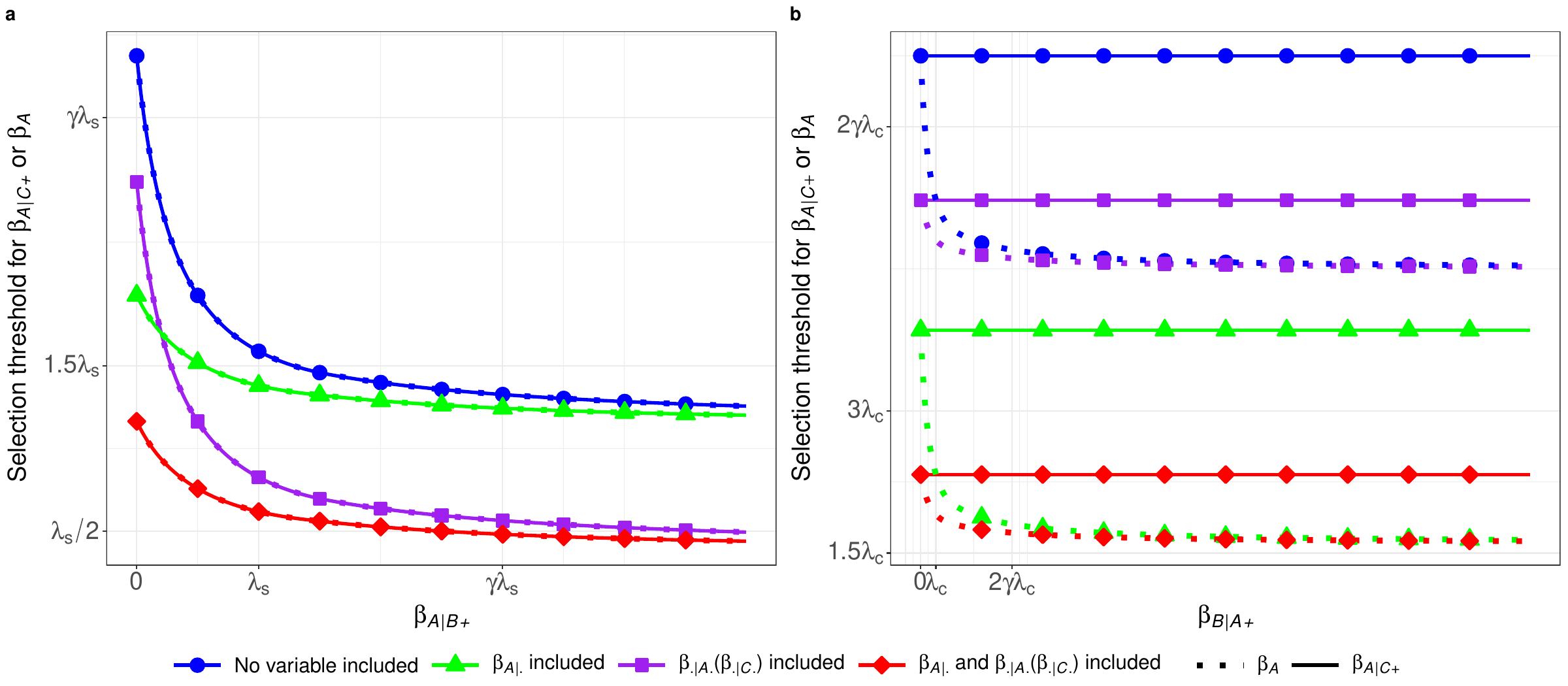}
\end{center}
\caption{\begin{revised} 
Effect of varying CME coefficients on selection thresholds of \textit{adaptive cmenet}: (a) Vary $\beta_{A|B+}$; (b) Vary $\beta_{B|A+}$. Curves show the threshold for $\beta_{A|C+}$ (dotted line) and $\beta_{A}$ (solid line) under four contexts: when no other variable included (blue), one sibling included (green), one cousin included (purple), and one sibling and one cousin included (red) in the model. Baseline adaptive cmenet setting: $(\lambda_s,\lambda_c,\gamma,\tau)=(1, 0.5, 3, 0.25)$. The adding variable has the same coefficient as $\beta=0.5$. \label{fig:3}
\end{revised}}
\end{figure}

In this section, we further investigate the coupling ability outlined in Section \ref{sec:problem}, focusing on the selection threshold  $\tilde{t}_{\beta_{j|k+}}$ for $J|K+$ after incorporating adaptive weights, derived from \eqref{equ5}:
\begin{equation}
\tilde{t}_{\beta_{j|k+}} = \lambda_{\mathcal{S}(j)}\exp\{-\frac{\tau \Vert \boldsymbol{\beta}_{\mathcal{S}(j)} \Vert_{\lambda_{\mathcal{S}(j)},\omega,\gamma}}{\lambda_{\mathcal{S}(j)}}\} + \lambda_{\mathcal{C}(k)}\exp\{-\frac{\tau \Vert \boldsymbol{\beta}_{\mathcal{C}(k)} \Vert_{\lambda_{\mathcal{C}(k)},\omega,\gamma}}{\lambda_{\mathcal{C}(k)}}\}.
\label{equ10}
\end{equation} 

\begin{revised}
In contrast to the fixed plateaus in Figure \ref{fig:1}, Figure \ref{fig:3} shows that the adaptive weights create lower and signal-dependent plateaus. This pattern comes from the adaptive weights: small coefficients in $\mathcal{S}(A)$ receive larger weights, so the threshold reacts strongly when the signal first appears. As coefficients in $\mathcal{S}(A)$ become larger, its weight shrinks and the penalty term approaches its plateau. This makes the threshold drop quickly at the start and then flatten earlier. Moreover, Figure \ref{fig:3} demonstrates a clearer form of \textit{CME coupling} principle: once a sibling or a cousin has been selected, the threshold shifts downward. A similar downward trend is observed for the dotted curves for $A$ in Figure \ref{fig:3}(b), while the solid curves for $A|C+$ remain flat as $\beta_{B|A+}$ increases, aligning with the \textit{CME reduction} principle.
The proposed method addresses both of the previous limitations: First, it dynamically ``rebalances" the between-group influence, which is proven in Figure \ref{fig:3}(a) by the crossover of the green and purple lines. At first, when the sibling signal is weak (near $\beta_{A|B+}=0$), the purple line is lower. However, as the sibling signal strengthens, the green line drops more steeply and crosses the purple line, proving the relative group influence is no longer fixed. Second, the proposed method successfully diminishes the within-group marginal effect. In Figure \ref{fig:3}(a), as $\beta_{A|B+}$ increases, the green line approaches the blue line at the higher plateau, and the red line approaches the purple line at the lower plateau. This indicates that, once the sibling group becomes strong, the extra sibling has a negligible marginal effect as its contribution is downweighted and the inner penalty is near saturation. Signals coming from the cousin side still matter, since the adjustments for cousins are not driven by how much activity we have in $\mathcal{S}(A)$. A similar observation can be seen for the dotted line in Figure \ref{fig:3}(b): the marginal effect of adding a new cousin diminishes as the cousin effect $\beta_{B|A+}$ increases, while adding a new sibling still matters.
\end{revised}

\section{Parameter Estimation}
\label{opt_algt}

For parameter estimation in GLM, the iteratively reweighted least squares (IRLS) algorithm is commonly used. 
The IRLS utilizes a second-order Taylor expansion to approximate the loss function quadratically around the current estimate \(\boldsymbol{\beta}^{(m)}\). 
The coefficients are then iteratively updated under the format of a weighted least squares as
$\frac{1}{2n}(\tilde{\mathbf{y}}-\mathbf{X}\boldsymbol{\beta})^{T}\mathbf{W}(\tilde{\mathbf{y}}-\mathbf{X}\boldsymbol{\beta})$, where
\(\tilde{\mathbf{y}} = \mathbf{X}\boldsymbol{\beta}^{(m)} + \mathbf{W}^{-1}(\mathbf{y} - \mathbf{v})\) represents the working response with \(\mathbf{v}\) and \(\mathbf{W}\) to be the first and second derivatives of \(L(\boldsymbol{\eta})\) with respect to \(\boldsymbol{\eta}\), evaluated at \(\boldsymbol{\eta}^{(m)}=\mathbf{X}\boldsymbol{\beta}^{(m)}\). Notably, \(\mathbf{W}\) is a diagonal matrix, where each diagonal entry corresponds to the weight of the \(i\)th observation under the least squares loss. At each iteration, the objective function with adaptive cmenet penalization can be expressed as:
\begin{equation}
Q(\boldsymbol{\beta})=\frac{1}{2n}(\tilde{\mathbf{y}}-\mathbf{X}\boldsymbol{\beta})^{T}\mathbf{W}(\tilde{\mathbf{y}}-\mathbf{X}\boldsymbol{\beta})+\tilde{P}_S(\boldsymbol{\beta})+\tilde{P}_C(\boldsymbol{\beta}).
\label{equ7}
\end{equation}
The $Q(\cdot)$ will exclusively refer to the formulation in \eqref{equ7} for the remainder of the paper.

To minimize the objective function in \eqref{equ7}, we employ the local coordinate descent (LCD) algorithm \citep{fu1998penalized, friedman2007pathwise, mazumder2011sparsenet} to efficiently derive a closed-form threshold operator of the coordinate-wise objective $Q_j(\cdot)$ with respect to coefficient $\beta_{j}$ while keeping the remaining coefficients fixed:
\begin{equation}
\begin{aligned}
Q_j(\boldsymbol{\beta})&=\frac{1}{2n}(\tilde{\mathbf{y}}-\mathbf{X}_{-j}\boldsymbol{\beta}_{-j}-\mathbf{x}_{j}^T\boldsymbol{\beta}_{j})^{T}\mathbf{W}(\tilde{\mathbf{y}}-\mathbf{X}_{-j}\boldsymbol{\beta}_{-j}-\mathbf{x}_{j}^T\boldsymbol{\beta}_{j}) \\ 
&+\eta_{\lambda_{\mathcal{S}(j)},\gamma}\{\Vert \boldsymbol{\beta}_{\mathcal{S}(j)} \Vert_{\lambda_{\mathcal{S}(j)},\omega,\gamma}\}+\eta_{\lambda_{\mathcal{C}(j)},\gamma}\{\Vert \boldsymbol{\beta}_{\mathcal{C}(j)} \Vert_{\lambda_{\mathcal{C}(j)},\omega,\gamma}\}.
\label{equ7.Qj}
\end{aligned}
\end{equation}
The coordinate-wise objective $Q_{j|k+}(\cdot)$ with respect to coefficient $\beta_{j|k+}$ is defined similarly:
\begin{equation}
\begin{aligned}
Q_{j|k+}(\boldsymbol{\beta})&=\frac{1}{2n}(\tilde{\mathbf{y}}-\mathbf{X}_{-j|k+}\boldsymbol{\beta}_{-j|k+}-\mathbf{x}_{j|k+}^T\boldsymbol{\beta}_{j|k+})^{T}\mathbf{W}(\tilde{\mathbf{y}}-\mathbf{X}_{-j|k+}\boldsymbol{\beta}_{-j|k+}-\mathbf{x}_{j|k+}^T\boldsymbol{\beta}_{j|k+}) \\ 
&+\eta_{\lambda_{\mathcal{S}(j)},\gamma}\{\Vert \boldsymbol{\beta}_{\mathcal{S}(j)} \Vert_{\lambda_{\mathcal{S}(j)},\omega,\gamma}\}+\eta_{\lambda_{\mathcal{C}(k)},\gamma}\{\Vert \boldsymbol{\beta}_{\mathcal{C}(k)} \Vert_{\lambda_{\mathcal{C}(k)},\omega,\gamma}\}.
\label{equ7.Qjk}
\end{aligned}
\end{equation}The LCD algorithm iteratively optimizes one coordinate at a time while holding the others fixed, offering computational efficiency and stability, particularly for high-dimensional problems. However, due to the presence of penalty terms, the objective function does not have a closed-form one-dimensional solution. To address this limitation, the majorization-minimization (MM) approach \citep{hunter2000quantile} is applied to construct a surrogate function for $Q_j(\cdot)$ or $Q_{j|k+}(\cdot)$, enabling the derivation of a closed-form threshold operator. 

\subsection{Convexity and Stability}

For parameter estimation, it is useful to examine the convexity property of the objective function in optimization.
For simplicity and clarity, we assume throughout the remainder of this paper that each column \(\mathbf{x}_j\) of \(\mathbf{X}\) is normalized, such that \(\mathbf{x}_j^T \mathbf{I}_n = 0\) and \(\frac{1}{n}\|\mathbf{x}_j\|_2^2 = 1\) for all \(j = 1, \dots, p'\). Unless otherwise specified, we denote all coordinate-wise objective functions in \eqref{equ7.Qj} and \eqref{equ7.Qjk} as $Q_j(\cdot)$, for $j=1,\dots,p'$ without differentiating between MEs and CMEs. 

\begin{proposition}
Define $\omega_{\max}=\max_{1\leq j \leq p'}\omega_j$ as the maximum individual weight. The objective function \(Q(\boldsymbol{\beta})\) in \eqref{equ7} is strictly convex whenever \(\tau + \frac{1}{\gamma \omega_{\max}} < \frac{\zeta_{\min}(\mathbf{X^T WX})}{2n\omega^2_{\max}}\), where \(\mathbf{W}\) is evaluated at \(\boldsymbol{\beta}\) and \(\zeta_{\min}\) is the minimum eigenvalue. 
\label{prop1}
\end{proposition}

Proposition \ref{prop1} implies that a small value of \(\tau + \frac{1}{\gamma \omega_{\max}}\) ensures the convexity of the objective function, leading to a global minimum. However, this condition is quite restrictive and typically applies only in low-dimensional settings where \(n \leq p'\). In high-dimensional scenarios, the coordinate descent algorithm remains effective, as it only requires local convexity for each coordinate. For linear regression with \(\mathbf{W = I}\), the local convexity of the coordinate-wise objective function \(Q_j(\beta_j)\) is guaranteed if \(\tau + \frac{1}{\gamma \omega_{\max}} \leq \frac{1}{2 \omega^2_{\max}}\). 

For generalized linear models, the situation becomes more complex due to the introduction of a non-identity weight matrix \(\mathbf{W}\), where \(\frac{1}{n}\mathbf{X^T WX} \neq \mathbf{I}\). Despite this, local convexity can still be ensured by imposing an upper bound on \(\mathbf{W}\), which corresponds to the Hessian matrix of the loss function with respect to the linear predictor \(\boldsymbol{\eta}\). Let \(\bar{w} = \max_i \sup_{\eta}\{\nabla^2 L_i(\eta)\}\), where \(L_i(\eta)\) is the loss function for the \(i\)th observation. For logistic regression, this bound is easily derived as \(\bar{w} = 1/4\), since \(w_i = \pi_i(1 - \pi_i)\). The following corollary provides some insights into maintaining local convexity and ensuring the effectiveness of the coordinate descent algorithm in high-dimensional settings for the logistic regression.

\begin{corollary}
For a coordinate-wise objective function $Q_j(\beta_j)$ of logistic regression, the $Q_j(\beta_j)$ is strictly convex for any $j=1,\dots,p'$, whenever $\tau+\frac{1}{\gamma \omega_{\max}} \leq \frac{1}{8 \omega^2_{\max}}$.
\label{coro1}
\end{corollary}

\begin{remark}
It is important to note that Corollary \ref{coro1} does not necessarily extend to other generalized linear models, as the Hessian matrices of loss functions derived from different exponential family distributions may lack a finite upper bound. However, this limitation is not prohibitive, as suitable values for \(\tau\) and \(\gamma\) can still be identified to ensure the convexity of the objective function. Alternatively, a pseudo-upper bound can be introduced by initializing \(\bar{w} \leftarrow \max_i\{\nabla^2 L_i(\eta_i^*)\}\) at the start of each iteration. 
\label{remark1}
\end{remark}

\subsection{Threshold Operators}
\label{sec:threshold}
To derive the closed-form threshold operator, we first define the majorizing function of \(Q_j(\beta_j)\) and \(Q_{j|k+}(\beta_{j|k+})\) through the first-order Taylor expansion of the adaptive penalties, as established in Lemma \ref{lemma1}.
\begin{lemma}
With the adaptive cmenet penalization defined in \eqref{equ4}, let $\tilde{\eta}_{\lambda_{\mathcal{G}},\tau}(\boldsymbol{\beta}_{\mathcal{G}}|\boldsymbol{\beta}_{\mathcal{G}}^*)$ be the first-order Taylor expansion of $\eta_{\lambda_{\mathcal{G}},\tau}(\boldsymbol{\beta}_{\mathcal{G}})$ at fixed $\boldsymbol{\beta}_{\mathcal{G}}^* \in \mathbb{R}^{|{\mathcal{G}}|}$, where ${\mathcal{G}}$ represent the sibling or cousin group of one pre-specified effect. Then,
\begin{equation}
    \tilde{\eta}_{\lambda_{\mathcal{G}},\tau}(\boldsymbol{\beta}_{\mathcal{G}}|\boldsymbol{\beta}_{\mathcal{G}}^*)=\eta_{\lambda_{\mathcal{G}},\gamma}\{\Vert \boldsymbol{\beta}^*_{\mathcal{G}} \Vert_{\lambda_{\mathcal{G}},\omega,\gamma}\}+\Delta^*_{\mathcal{G}}\times\{\Vert \boldsymbol{\beta}_{\mathcal{G}} \Vert_{\lambda_{\mathcal{G}},\omega,\gamma}-\Vert \boldsymbol{\beta}^*_{\mathcal{G}} \Vert_{\lambda_{\mathcal{G}},\omega,\gamma}\}
\label{eta_1taylor}
\end{equation}
is a majorizing function of $\eta_{\lambda_{\mathcal{G}},\tau}(\boldsymbol{\beta}_{\mathcal{G}})$, where $\Vert \boldsymbol{\beta}_{\mathcal{G}}\Vert_{\lambda_\mathcal{G}, \omega,\gamma} := \sum_{l\in\mathcal{G}}\omega_lg_{\lambda_{\mathcal{G}},\gamma}(\beta_l)$. Furthermore, we can derive $\tilde{Q}_{j|k+}(\beta_{j|k+}|\boldsymbol{\beta}^*)$ as the majorizing function of $Q_{j|k+}(\beta_{j|k+})$ at fixed $\boldsymbol{\beta}^* \in \mathbb{R}^{p'}$,
\begin{equation}
\begin{aligned}
    \tilde{Q}_{j|k+}(\beta_{j|k+}|\boldsymbol{\beta}^*)&=L(\beta_{j|k+}|\boldsymbol{\beta}^*)+\eta_{\lambda_{\mathcal{S}(j)},\tau}(\Vert \boldsymbol{\beta}^*_{\mathcal{S}(j)} \Vert_{\lambda_{\mathcal{S}(j)},\omega,\gamma})+\eta_{\lambda_{\mathcal{C}(k)},\tau}(\Vert \boldsymbol{\beta}^*_{\mathcal{C}(k)} \Vert_{\lambda_{\mathcal{C}(k)},\omega,\gamma})\\
    &+\Delta^*_{\mathcal{S}(j)}\omega_{j|k+}\{g_{\lambda_{\mathcal{S}(j)},\gamma}(\beta_{j|k+})-g_{\lambda_{\mathcal{S}(j)},\gamma}(\beta^*_{j|k+})\} \\
    &+\Delta^*_{\mathcal{C}(k)}\omega_{j|k+}\{g_{\lambda_{\mathcal{C}(k)},\gamma}(\beta_{j|k+})-g_{\lambda_{\mathcal{C}(k)},\gamma}(\beta^*_{j|k+})\},
\end{aligned}
\label{equ8}
\end{equation}
where $*$ indicates the quantity is computed with the current (i.e., most recently updated) value of $\boldsymbol{\beta}^*$. The majorizing function $\tilde{Q}_j(\beta_j|\boldsymbol{\beta}^*)$ for $Q_j(\beta_j)$ is defined in the similar manner. 
\label{lemma1}
\end{lemma}
For a given coordinate, minimizing the objective function \(Q_j(\beta_j)\) is equivalent to regressing the partial residuals \(\mathbf{y} - \mathbf{X}_{-j}\boldsymbol{\beta}^*_{-j}\) on \(\mathbf{x}_j\). Expressing the gradient of \(L(\boldsymbol{\beta})\) with respect to \(\beta_j\), while holding other coefficients fixed, yields \(\nabla L(\beta_j | \boldsymbol{\beta}^*) = -z_j + v_j \beta_j\), where \(z_j = \frac{1}{n} \mathbf{x}_j^T \mathbf{W}(\tilde{\mathbf{y}} - \mathbf{X}_{-j}\boldsymbol{\beta}^*_{-j})\) represents the unpenalized solution for \(\beta_j\), and \(v_j = \frac{1}{n} \mathbf{x}_j^T \mathbf{W} \mathbf{x}_j\). Therefore, the coordinate-wise update for \(\beta_j\) (or \(\beta_{j|k+}\)) in generalized linear regression with adaptive cmenet penalization relies on a closed-form threshold function with respect to \(z_j\) (or \(z_{j|k+}\)), expressed as:
\begin{equation*}
    \begin{aligned}
        \beta_j &\leftarrow S_{\lambda_{\mathcal{S}(j)},\lambda_{\mathcal{C}(j)}}(z_j;\Delta_{\mathcal{S}(j)},\Delta_{\mathcal{C}(j)}),\\
        \text{or} \quad \beta_{j|k+} &\leftarrow S_{\lambda_{\mathcal{S}(j)},\lambda_{\mathcal{C}(k)}}(z_{j|k+};\Delta_{\mathcal{S}(j)},\Delta_{\mathcal{C}(k)}).
    \end{aligned}
\end{equation*}
where \(S(\cdot; \Delta_1, \Delta_2)\) is defined in the following Theorem \ref{thero1}.

\begin{theorem}
Suppose $\tau+\frac{1}{\gamma \omega_{\max}} \leq \frac{1}{8 \omega^2_{\max}}$, then the threshold function used to update a regression coefficient for \eqref{equ7} is defined as: 
\begin{equation}
\begin{aligned}
&S_{\lambda_1,\lambda_2}(z_j;\Delta_1,\Delta_2) = \\
&\left\{
    \begin{aligned}
        & \frac{z_j}{v_j},  &&\mathrm{if}\quad|z_j|\in[v_j\gamma\lambda_{(1),\infty}),\\
        &\frac{\mathrm{sgn}(z_j)(|z_j|-\Delta_{(1)}\omega_j)}{v_j-\frac{\Delta_{(1)}\omega_j}{\lambda_{(1)}\gamma}}, &&\mathrm{if}\quad|z_j|\in[v_j\lambda_{(2)}\gamma+\Delta_{(1)}\omega_j(1-\frac{\lambda_{(2)}}{\lambda_{(1)}}),v_j\lambda_{(1)}\gamma),\\
        &\frac{\mathrm{sgn}(z_j)(|z_j|-\Delta_{(1)}\omega_j-\Delta_{(2)}\omega_j)}{v_j-\frac{\Delta_{(1)}\omega_j}{\lambda_{(1)}\gamma}-\frac{\Delta_{(2)}\omega_j}{\lambda_{(2)}\gamma}}, &&\mathrm{if}\quad|z_j|\in[\Delta_{(1)}\omega_j+\Delta_{(2)}\omega_j,v_j\lambda_{(2)}\gamma+\Delta_{(1)}\omega_j(1-\frac{\lambda_{(2)}}{\lambda_{(1)}})),\\
        &0, &&\mathrm{if}\quad |z_j|\in [0,\Delta_{(1)}\omega_j+\Delta_{(2)}\omega_j).
    \end{aligned}
     \right.
\end{aligned}
\label{equ9}
\end{equation}
\label{thero1}
where $\lambda_{(1)}=\max(\lambda_{1},\lambda_{2})$ and $\lambda_{(2)}=\min(\lambda_{1},\lambda_{2})$, with $\Delta_{(1)}$ and $\Delta_{(2)}$ its corresponding slopes.
\end{theorem}

\begin{figure}[t]
\begin{center}
\includegraphics[width=1\textwidth]{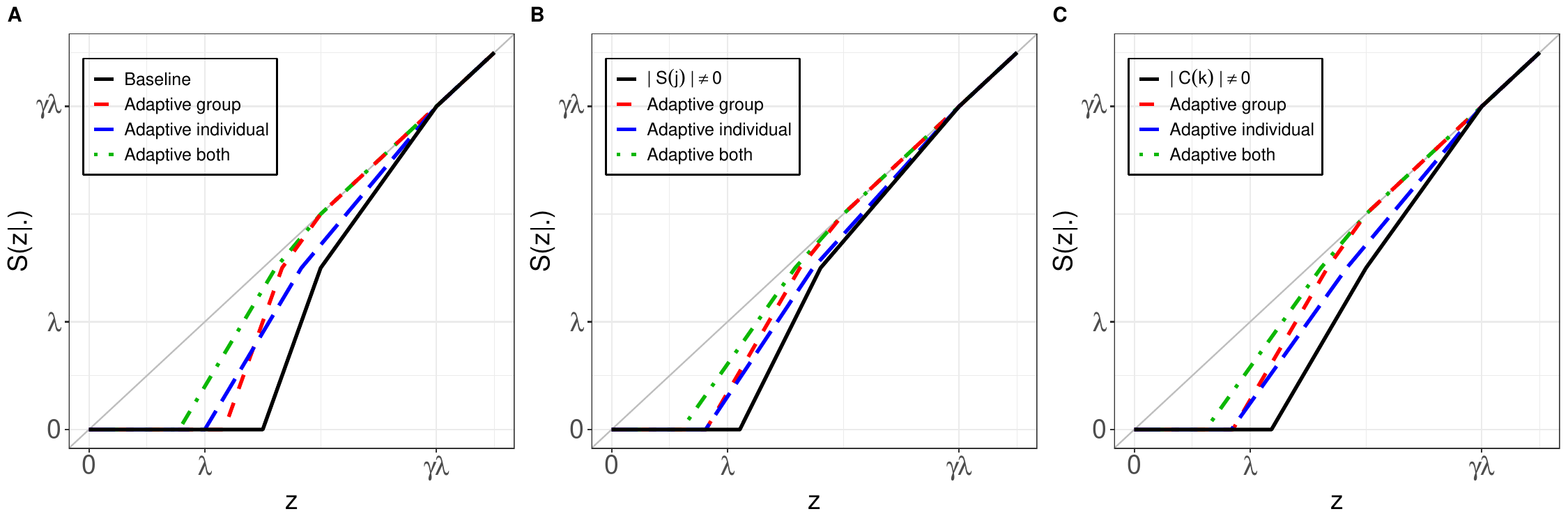}
\end{center}
\caption{Comparison of the threshold function $S(\cdot; \Delta_1, \Delta_2)$ between \textit{cmenet} and \textit{adaptive cmenet} across three scenarios: panel (A) No variable included, panel (B) inclusion of one sibling with coefficient $\beta=2$, and panel (C) inclusion of one cousin with coefficient $\beta=2$. Each plot illustrates four threshold functions: (solid) unweighted penalty, i.e., baseline cmenet with settings $(\lambda_s, \lambda_c, \gamma, \tau) = (1, 0.5, 3, 0.25)$; (dash) group-weighted penalty with adaptive weights $(\Omega_{\mathcal{S}(j)}, \Omega_{\mathcal{C}(j)}, \omega_j) = (2/3, 1, 1)$; (long dash) individual-weighted penalty with adaptive weights $(\Omega_{\mathcal{S}(j)}, \Omega_{\mathcal{C}(j)}, \omega_j) = (1, 1, 1.5)$; and (dot dash) combined group and individual-weighted penalty with adaptive weights $(\Omega_{\mathcal{S}(j)}, \Omega_{\mathcal{C}(j)}, \omega_j) = (2/3, 1, 1.5)$. \label{fig:2}}
\end{figure}

It is critical to understand how group and individual weights will influence the threshold function \(S(\cdot; \Delta_1, \Delta_2)\). 
As illustrated in Figure \ref{fig:2}, the threshold function demonstrates how changes in penalty weights affect both the turning point and the slope of the curve. 
Introducing group weights into the penalty terms, represented by the dashed line across the three plots, shifts the turning point of the threshold function to the left. This shift indicates that the model becomes more likely to include effects at a lower selection threshold. 
Additionally, incorporating individual weights, shown by the long dashed line, modifies the slope of the threshold function. This adjustment reduces the shrinkage rate of the coefficients, enabling a more gradual penalization. 
When both group and individual weights are applied, as depicted by the dot dashed line, the resulting threshold function is the least penalized. This configuration aligns more closely with the diagonal line, reflecting a generalized and less restrictive approach to penalization. These results highlight the flexibility of the adaptive cmenet penalty structure. Specifically, group weights primarily adjust the selection threshold for effect inclusion, while individual weights regulate the rate of shrinkage. 
By combining these two weighting schemes, the proposed framework provides a more flexible and adaptive approach to variable selection.

\subsection{The \texttt{glmcmenet} Algorithm and Parameter Tuning}

\begin{revised} 
The detailed algorithm of \texttt{glmcmenet} using the threshold function is available in the supplemental materials. For the parameter tuning, we use K-fold cross-validation to find an optimal set of parameters $(\lambda_s, \lambda_c, \gamma, \tau)$. To efficiently explore the tuning parameter space, we adopt the same strategy as \cite{mak2019cmenet}. Specifically, we first scan a log-spaced grid for the structure parameters $(\gamma, \tau)$ and pick the best pair by cross-validation (CV); conditional on that choice, we search a log-spaced grid for the penalty magnitudes $(\gamma, \tau)$ pair, and then a log-spaced grid for the penalty magnitudes $(\lambda_s, \lambda_c)$. The penalties $\lambda_s$ and $\lambda_c$ are global, while adaptivity is introduced via sibling and cousin group weights $\Omega_{\mathcal{S}(j)}$, $\Omega_{\mathcal{C}(j)}$, and an individual weight $w_j$ for effect $J$. To avoid trivial solutions and reduce the search space, we use a weighted start rule: with standardized $\mathbf{X}$ and centered $\mathbf{y}$, define
$$\lambda_{\max}^{\omega}(\rho)=\max_{j}\frac{|\mathbf{x}^T_j\mathbf{y}|/n}{\rho\Omega_{\mathcal{S}(j)wj}+(1-\rho)\Omega_{\mathcal{C}(j)w_j}},$$
where $\rho=\lambda_s/(\lambda_s+\lambda_c) \in (0,1)$. Then the all-zero solution satisfies the KKT conditions (i.e., is a stationary solution) whenever $\lambda_s+\lambda_c \geq \lambda_{\max}^{\omega}$. Accordingly, for each fixed 
$\rho$ we restrict the grid to $\lambda_s+\lambda_c \geq \lambda_{\max}^{\omega}(\rho)$, and set $(\lambda_s,\lambda_c)=(\rho(\lambda_s+\lambda_c),(1-\rho)(\lambda_s+\lambda_c))$.

For GLMs with canonical links $g(\mu)=\eta$, we use the analogous weighted start rule $\lambda_{\max}^{\omega}=\max_j\frac{| \mathbf{x}^T_j\boldsymbol{u} |/n}{\rho\Omega_{\mathcal{S}(j)wj}+(1-\rho)\Omega_{\mathcal{C}(j)w_j}}$ with $\boldsymbol{u}=\frac{\mathbf{y}-\boldsymbol{\mu}}{\text{Var}(\mathbf{Y}|\boldsymbol{\eta})}\cdot \frac{\partial \boldsymbol{\mu}}{\partial \boldsymbol{\eta}}$. Moreover, if $Q(\boldsymbol{\beta})$ is strictly convex, it is the unique minimizer based on Proposition \ref{prop1}. Therefore, we also enforce the stability condition to guarantee the strict convexity of the coordinate-wise subproblem. For computational efficiency during CV, we leverage the same techniques as those employed in cmenet, including warm starts and active set optimization as \cite{mak2019cmenet}, thus the same stability guarantees and practical behavior apply.
\end{revised} 

One notable advantage of the developed algorithm is the guaranteed descent property due to the MM framework, combined with the cycling procedure of LCD. 
These techniques ensure that the objective function decreases monotonically at each iteration. Moreover, by leveraging the local convexity property, we can establish that the algorithm converges to a stationary point, providing enhanced stability to the optimization process.

\begin{proposition}
    let $\{\boldsymbol{\beta}^{(m)}\}$ denote the sequence of coefficient estimates generated at each iteration of the proposed algorithm for logistic regression with adaptive cmenet penalization. For all iterations $m=0,1,2,\dots$, the following inequality holds:
    \begin{equation*}
    Q(\boldsymbol{\beta}^{(m+1)})\leq Q(\boldsymbol{\beta}^{(m)}).
    \end{equation*}
    Furthermore, when $\tau+\frac{1}{\gamma \omega_{\max}} \leq \frac{1}{8 \omega^2_{\max}}$, the algorithm is guaranteed to converge to a stationary point of $Q(\boldsymbol{\beta})$.
    \label{prop2}
\end{proposition}

\section{Simulation Study}
\label{sec:sim}
In this section, two scenarios are considered to examine different aspects of the proposed method.
In the first scenario (Section \ref{sim:first}), we focus on continuous responses, using the exact same simulation settings as those used in the cmenet paper \citep{mak2019cmenet}. 
Here, the active effects consist entirely of pure siblings, pure cousins, or pure main effects. 
The objective of this scenario is to assess the effectiveness of adding adaptive weights compared to the original cmenet method. 
This scenario serves as a controlled experiment to demonstrate the improvements because of adaptive weights. 
In the second scenario (Section \ref{sim:second}), we consider mixed effects where the active effects include combinations of main effects and their associated siblings or cousins. It reflects more realistic applications. This scenario evaluates the performance of the proposed for both continuous and binary responses, providing a broader assessment of its effectiveness across different data types. 
Each simulation case in both scenarios is replicated 100 times to ensure the robustness of the results.

\begin{figure}[t]
\begin{center}
\includegraphics[width=1\textwidth]{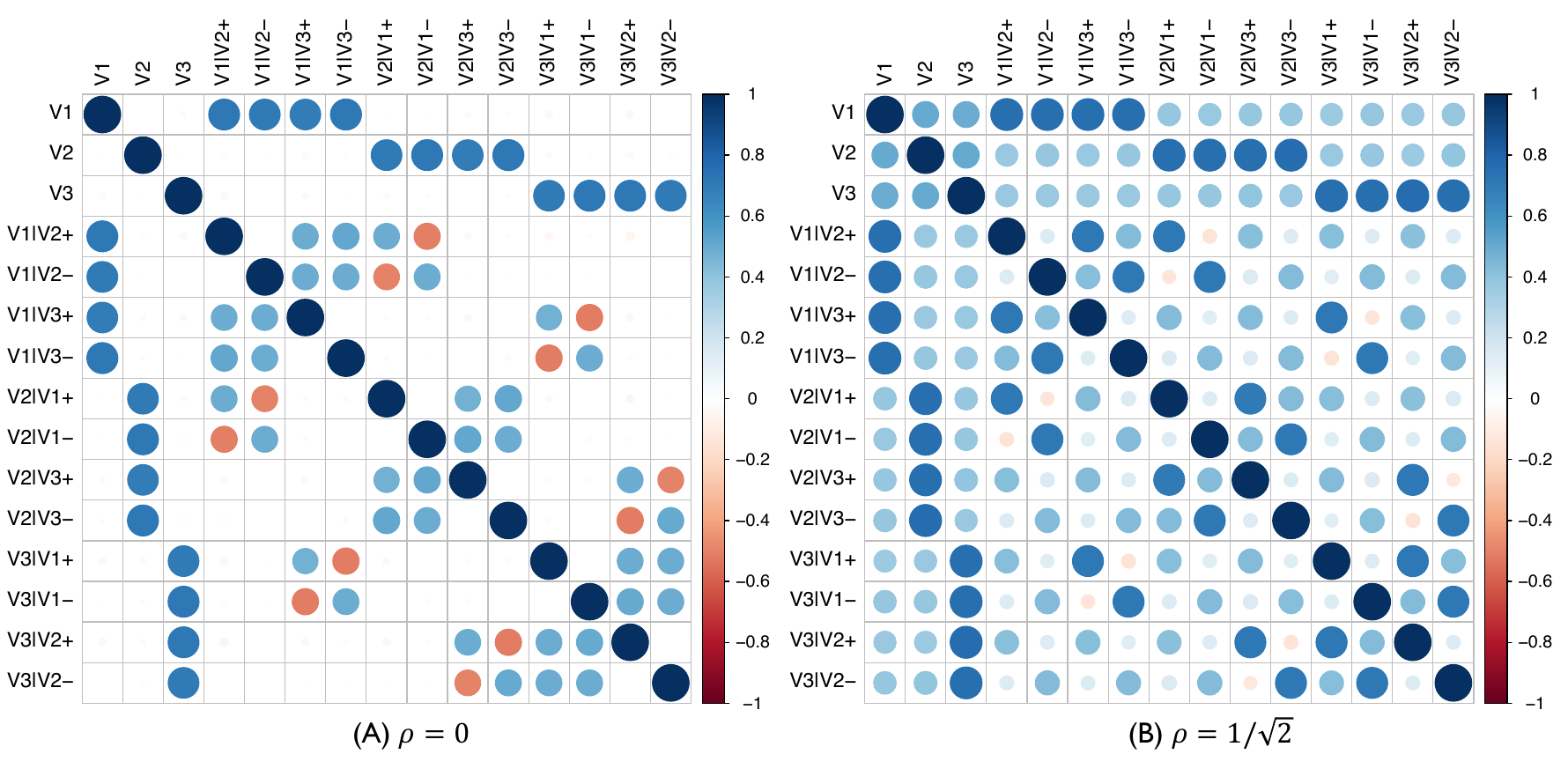}
\end{center}
\caption{Correlation matrix of the simulated full model matrix ($n=100$, $p=3$) generated from the equicorrelation latent model with (A) $\rho=0$ and (B) $\rho=1/\sqrt{2}$.}
\label{fig:corr}
\end{figure}

The model matrix $\mathbf{X}$ is generated using the equicorrelated latent model described in \cite{mak2019cmenet}. This approach constructs the MEs $\{\mathbf{x}_j\}_{j=1}^p \in \{-1,+1\}^n$ from a latent matrix $\mathbf{Z}={(z_{i,j})_{i=1}^{n}}_{j=1}^p \in \mathbb{R}^{n\times p}$, where each row follows a multivariate normal distribution $\mathbf{z}_i\sim \mathcal{N}\{\mathbf{0},\rho \mathbf{J}_p+(1-\rho)\mathbf{I}_p\}$ ($\mathbf{J}_p$ is the $p \times p$ matrix of one, $\mathbf{I}_p$ is the $p \times p$ identity matrix, and $\rho \in [0,1]$ is the correlation parameter). The binary MEs are subsequently derived from this latent structure:
\begin{equation} 
x_{i,j}=-1\{z_{i,j}\leq0\}+1\{z_{ij}>0\}, \quad i=1,\dots.n, j=1,\dots,p.
\label{equ:latent_model}
\end{equation}
Using the equicorrelated latent model, it is straightforward to construct $4{p \choose 2}$ CMEs based on the definition provided in Section \ref{sec:cme}. As a result, the full model matrix for each simulation is represented as $\mathbf{X}=\{\mathbf{x}_i\}_{i=1}^{n}\in \mathbb{R}^{n\times p'}$, where each row vector is defined as $\mathbf{x}_i=(x_{i,1},\dots,x_{i,p},x_{i,p+1},\dots,x_{i,p'})'\in \mathbb{R}^{p'\times 1}$. The correlation structure of this design matrix for the case where $p=3$ is illustrated in Figure \ref{fig:corr} (see \cite{mak2019cmenet}, Section 2.2, for additional details).

Two different types of response variables are considered: continuous and binary.
Here we focus on logistic regression as a representative case, but the methodology applies similarly to other GLMs.
The continuous response variable $y_i$ variable in this study is generated by $y_i=\beta_0+\sum_{j=1}^{p'}\beta_j x_{i,j}+\epsilon_i$, where $\epsilon_i \sim N(0,1)$. The binary response is generated based on a logistic regression model. The linear predictor, $\eta_i$, is calculated as $\eta_i = \beta_0+\sum_{j=1}^{p'}\beta_j x_{i,j}$. Then the binary response variable $y_i$ is simulated from a Bernoulli distribution
$ y_i \sim \text{Bernoulli}\left(\frac{1}{1 + \exp(-\eta_i)}\right)$.

The \texttt{cmenet} (R package \texttt{cmenet} V0.1.2; \citealp{cmenet}) is compared to the proposed approach to assess the effectiveness of adaptive weights. Several benchmark methods of variable selection are also considered, including the Lasso (R package \texttt{glmnet} V4.1.8; \citealp{glmnet,glmnet_glm}), the adaptive Lasso (R package \texttt{glmnet} V4.1.8; \citealp{zou2006adaptive}), the non-concave penalty method MCP (R package \texttt{ncvreg} V3.14.3; \citealp{ncvreg, zhang2010nearly}), the bi-level selection method GEL (R package \texttt{grpreg} V3.5.0; \citealp{breheny2015group}), and the popular two-way interaction selection method \texttt{hierNet} (R package \texttt{hierNet} V1.9; \citealp{hierNet}). Three criteria are used for measuring the selection accuracy: $F_1$ score, Precision, and True Positive Rate (TPR) or Recall. Denote the number of true positives, false positives, true negatives, and false negatives as TP, FP, TN, and FN, respectively. We can calculate the Precision as Precision=TP/(TP+FP), the Recall as Recall=TP/(TP+FN), and the $F_1$ score as 
$F_1=(2\text{Precision}\times \text{Recall})/(\text{Precision}+\text{Recall})$. In addition to evaluating selection performance, we also measure the prediction performance. The metrics for prediction evaluation depend on the type of response. For continuous responses, we evaluate the performance using the mean squared prediction error (MSPE). For binary responses, we assess performance based on the misclassification rate (MCR). Here, $\text{MSPE} =  \frac{1}{n}\sum_{i=1}^n(y_{i,new}-\hat{y}_{i,new})^2,$ and $
\text{MCR} = \frac{1}{n}\sum_{i=1}^n I(y_{i,new} \neq \hat{y}_{i,new})$. \begin{revised} Since \texttt{hierNet} does not encode conditional relationships within interactions, element-wise selection metrics are not directly comparable. Therefore, for \texttt{hierNet} we report model size (number of selected effects) and out-of-sample prediction error as indicators of parsimony and predictive performance.
According to the principle of parsimony \citep{stoica1982parsimony, wu2011experiments}, when two models achieve comparable prediction performance, the model with fewer parameters is preferable to one with a larger number of parameters. \end{revised} 

\subsection{Evaluation on adaptive weights}
\label{sim:first}

In this section, we assess the performance of adaptive cmenet under the same simulation settings as those used for cmenet in \citet{mak2019cmenet}, with focus exclusively on continuous responses. Following their framework, we consider three distinct types of active effects in the underlying true model: (1) all active effects are main effects, such as \(\{A, B, C, D\}\); (2) all active effects are sibling effects, such as \(\{A|B+, A|D+, B|C-, B|E+\}\); and (3) all active effects are cousin effects, such as \(\{C|A+, B|A-, D|B+, E|B+\}\). For the cases involving sibling and cousin effects, we ensure that no other types of group structures are included among the active effects. In this study, the sample size is \(n=100\), and the number of MEs is \(p=40\), resulting in a total of \(p' = p + 4{p \choose 2} = 3160\) effects (including both MEs and CMEs). The coefficients for active effects are set to \(\beta_j = 5\) and the intercept is \(\beta_0=12\). To evaluate the impact of correlations among MEs, this study considers two correlation settings: uncorrelated (\(\rho=0\)) and moderately correlated (\(\rho=1/\sqrt2\)). The number of active groups is varied (\(\{4,6,8,10,12\}\)), with two active effects within each group.

\begin{sidewaysfigure}
    \centering
     \includegraphics[width=1\textwidth]{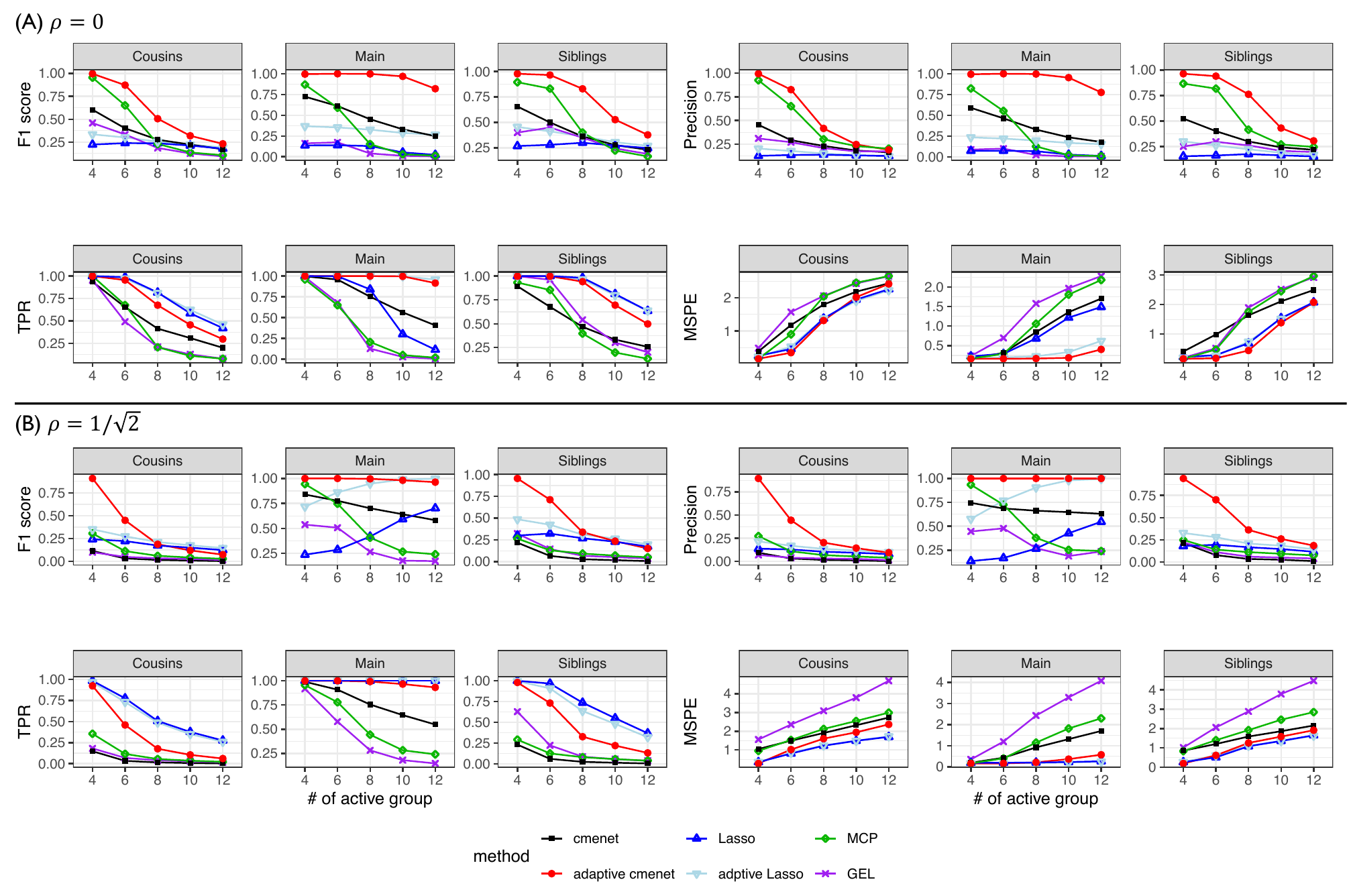}   
    \caption{Example 1. Simulation results for continuous responses with $n=100$, $p=40$ for (A)uncorrelated MEs $\rho=0$ and (B) correlated MEs $\rho=1/\sqrt{2}$. The x-axis shows the number of active groups, while the y-axis displays the $F_1$ score, Precision, True Positive Rate (TPR), and MSPE. Each column corresponds to a type of active effect: cousins (left), main effects (middle), and siblings (right).}
    \label{fig:gauss_res}
\end{sidewaysfigure}

Figure \ref{fig:gauss_res} shows the simulation results for scenario 1.  Across nearly all simulation settings of the uncorrelated case, the proposed method performs significantly better than not only the cmenet but also other well-established methods. By incorporating adaptive weights, it achieves the highest $F_1$ scores, striking a better balance between Precision and TPR. The adaptive cmenet always demonstrates much higher Precision by selecting fewer irrelevant variables, while maintaining competitive TPR. It also produces lower or comparable MSPE to other methods, highlighting its capability to deliver strong prediction performance alongside improved selection accuracy. 

Considering the adaptive cmenet to the Lasso and the adaptive Lasso, a trade-off becomes apparent. Sometimes, the Lasso-based methods demonstrate higher TPR than adaptive cmenet, which means they identify more true active effects. However, the adaptive cmenet achieves higher Precision, highlighting its ability to select fewer irrelevant variables. This pattern is persistent across cousins, main effects, and siblings, emphasizing the relative strengths of adaptive cmenet in maximizing true positive detection while producing parsimonious models. 

When comparing the relative performance across the three types of active effects, there are notable differences. Main effects are the easiest to identify, as indicated by consistently high selection criteria values, with many approaching or reaching 1. This suggests a near-perfect selection of all active effects without introducing false positives. In contrast, cousin effects are more difficult to detect, as evidenced by lower selection criteria values, particularly as the number of active groups increases. Sibling effects fall between these two extremes, exhibiting moderate performance in selection accuracy. The superior identification of main effects compared to cousin and sibling effects highlights the effectiveness of the CME reduction principle within the penalization framework.

Now considering the MEs are correlated ($\rho=1/\sqrt{2}$) as shown in Figure \ref{fig:gauss_res}(B), the proposed adaptive cmenet demonstrates a clear advantage over other methods in terms of selection accuracy, particularly when the number of active groups is small. The improvement in performance under correlated settings is more substantial than in the uncorrelated case. For instance, in the uncorrelated scenario, certain methods, such as the MCP, achieve comparable $F_1$ scores to adaptive cmenet for cousin and sibling effects. However, in the correlated case, the gap of $F_1$ scores between the adaptive cmenet and the MCP becomes more pronounced compared to the uncorrelated setting. This superiority only diminishes when the number of active groups is large. The result in the correlated case underscores the effectiveness of the CME coupling principle in capturing group structures within CMEs, as the CME group structure becomes more evident with higher correlations (Figure \ref{fig:corr}). 

\subsection{Evaluation on Complex Effect Structures}
\label{sim:second}

In the second simulation scenario, we consider two distinct types of active effects in the underlying true model for both continuous and binary response: (1) active effects consist of main effects and their siblings, such as $\{A, B, A|B+, A|D+, B|C-, B|E+\}$; and (2) active effects consist of main effects and their cousins, such as $\{A, B, C|A+, B|A-, D|B+, E|B+\}$. The inclusion of main effects introduces overlapping and ambiguous groupings. Unlike the first example, we do not ensure the absence of other types of group structures among the active effects. This setup, which closely resembles real-world scenarios, provides additional challenges for selection methods. Here, the sample size is \(n=50\), and the number of MEs is \(p=20\), leading to a total of \(p' = p + 4{p \choose 2} = 780\) effects. For this example, the coefficients of MEs are set to \(\beta_{\text{ME}} = 5\), while the coefficients of CMEs are \(\beta_{\text{CME}} = 1\). The correlation settings and the number of active groups follow the same configurations as in Scenario 1. \begin{revised} As a sensitivity analysis, the supplemental materials reports an additional experiment with a fixed total of four groups and exactly two active effects, where the effect size $\beta_{\text{CME}}$ varies from 0.25 to 16 (with fixed $\beta_{\text{CME}}$=5) to evaluate the impact of signal-to-noise ratio and effect-size ratio $\beta_{\text{ME}}/\beta_{\text{CME}}$ on selection and prediction.\end{revised}

\begin{sidewaysfigure}
    \centering
     \includegraphics[width=1\textwidth]{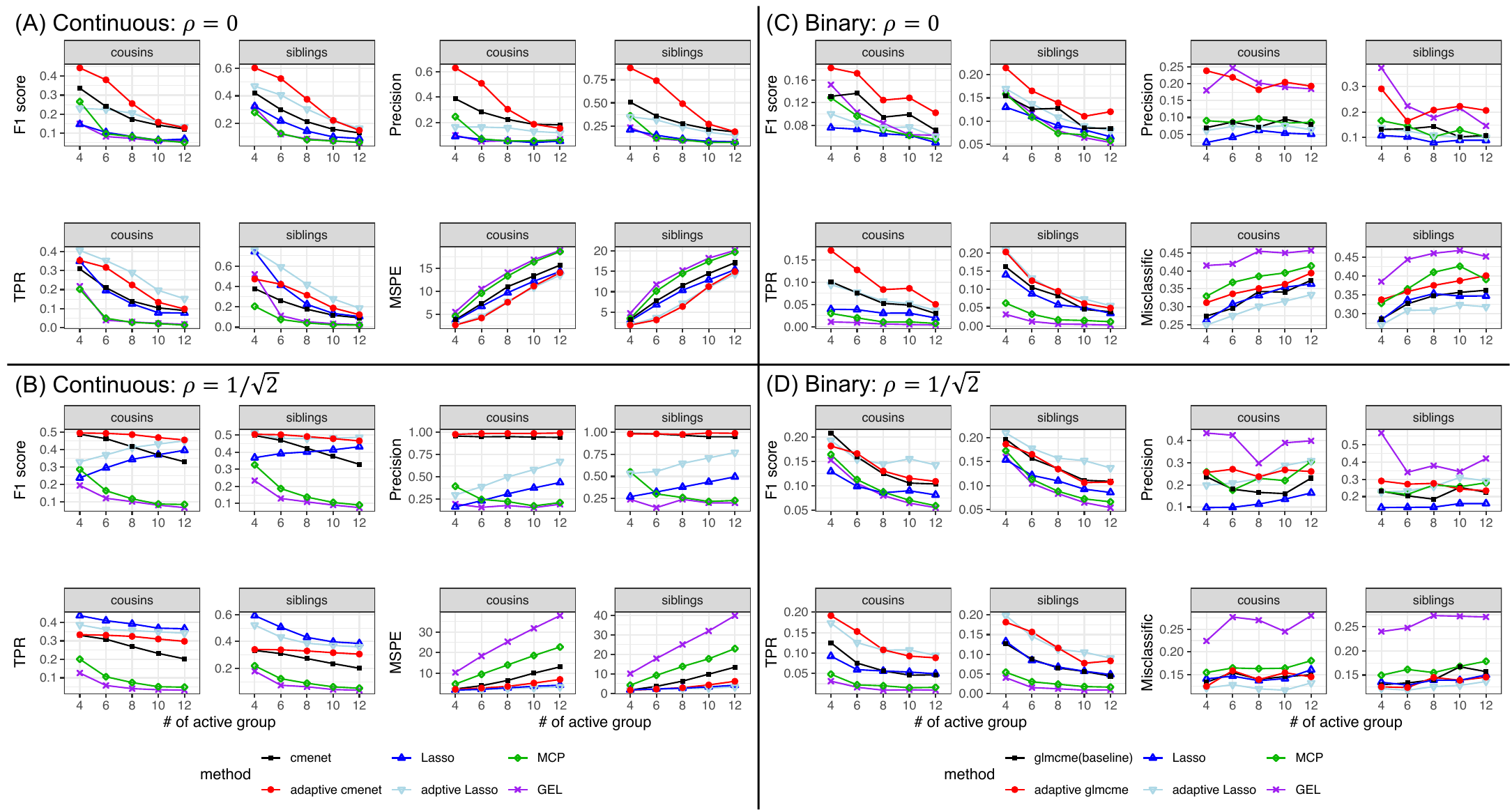}   
    \caption{Example 2. Simulation results with $n=50$, $p=20$ for (A) continuous response with uncorrelated MEs $\rho=0$, (B) continuous response with correlated MEs $\rho=1/\sqrt{2}$, (C) binary response with uncorrelated MEs $\rho=0$, and (D) binary response with correlated MEs $\rho=1/\sqrt{2}$. The x-axis shows the number of active groups, while the y-axis displays the $F_1$ score, Precision, True Positive Rate (TPR), and MSPE (for continuous) or Misclassification Rate (for binary). Each column corresponds to a type of active effect: main+cousins (left), and main+siblings (right).}
    \label{fig:mix_res}
\end{sidewaysfigure}

Considering continuous responses in Figure \ref{fig:mix_res}(A) and \ref{fig:mix_res}(B), the adaptive cmenet method outperforms other methods across both uncorrelated and correlated settings for main+cousin effects. The improvement is particularly clear in the correlated case, where the stronger group dependencies enhance adaptive cmenet’s ability to balance true positive and false positive rates. For main+sibling effects, the adaptive cmenet performs better than other methods in the uncorrelated case, reflecting the effectiveness of detecting sibling effects when group structures are weakly defined. In the correlated case, the adaptive cmenet demonstrates improved performance compared to the uncorrelated case, particularly as the number of active groups increases. However, the $F_1$ score curves of adaptive cmenet and adaptive Lasso largely overlap, indicating comparable overall performance, although the adaptive cmenet achieves higher Precision and lower TPR. A possible explanation is that both methods rely on the same ridge estimator for generating initial weights, which influences the penalty structure in both approaches. Across all scenarios, the adaptive cmenet consistently delivers the lowest MSPE, underscoring its ability to integrate accurate selection with strong predictive performance.

For the results of binary responses shown in Figure \ref{fig:mix_res}, the adaptive cmenet exhibits competitive performance across uncorrelated and correlated settings, effectively balancing selection accuracy and prediction. 
For main+cousin effects, the adaptive cmenet achieves the highest $F_1$ scores and TPR in the uncorrelated setting, highlighting its ability to capture true positive relationships while maintaining strong precision. 
For main+sibling effects, the adaptive cmenet performs competitively with the adaptive Lasso in both uncorrelated and correlated cases, offering a strong trade-off between the precision and TPR. 
While it does not surpass the adaptive Lasso in $F_1$ scores in some scenarios, it still maintains lower false discovery rates. Although the MCP and GEL methods exhibit slightly higher Precision when $\rho=1/\sqrt{2}$, the adaptive cmenet strikes a better balance overall by effectively managing both false discoveries and true positives. 

\begin{figure}
    \centering
     \includegraphics[width=1\textwidth]{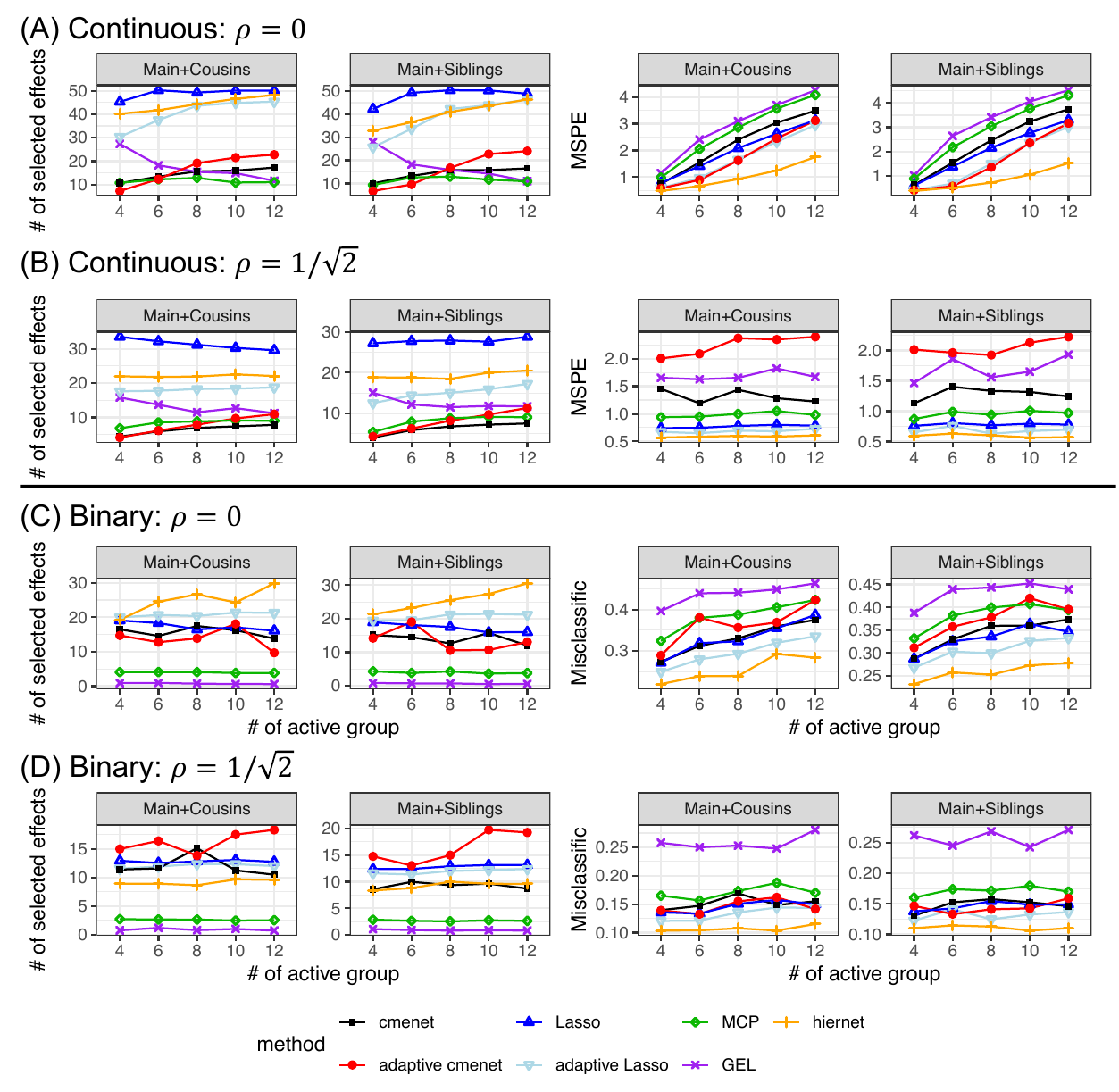}   
    \caption{\begin{revised}Example 2. Comparison to hierNet with $n=50$, $p=20$ for (A) continuous response with uncorrelated MEs $\rho=0$, (B) continuous response with correlated MEs $\rho=1/\sqrt{2}$, (C) binary response with uncorrelated MEs $\rho=0$, and (D) binary response with correlated MEs $\rho=1/\sqrt{2}$. The x-axis shows the number of active groups, while the y-axis displays the total \# of selected effects, and MSPE (for continuous) or Misclassification Rate (for binary). Each column corresponds to a type of active effect: main+cousins (left), and main+siblings (right).\end{revised}}
    \label{fig:exp2_hierNet}
\end{figure}

\begin{revised} In Figure \ref{fig:exp2_hierNet}, \texttt{hierNet} achieves relatively better predictive accuracy than adaptive cmenet with the advantage widening under correlation ($\rho=1/\sqrt2$) and as the number of active groups increases. However, the trade-off is obvious: \texttt{hierNet} selects noticeably larger models, suggesting it leverages additional pairwise interactions to capture dependence structure and boost prediction. By contrast, the adaptive cmenet produces more parsimonious models with an interpretable CME structure, making it preferable when interpretability and parsimony are the primary objectives. \end{revised}

\section{Case Study of Gene Association}
\label{sec:gene}

In this section, we apply the proposed method to two gene association studies to evaluate its performance in real-world datasets: one with a continuous response and the other with a binary response. In both studies, the main effects are binary single-nucleotide polymorphism (SNP) markers. Each SNP represents a specific genetic variation at a particular location in the genome and can exist in three genotypic forms: homozygous reference (AA), heterozygous (Aa), or homozygous variant (aa). Traditionally, SNPs are encoded using an additive $\{0,1,2\}$ coding scheme to represent AA, Aa, and aa, respectively \citep{lewis2002genetic}. In both studies, we analyze datasets from inbred lines, where the inbreeding process ensures that all single nucleotide polymorphisms (SNPs) are fully homozygous (either AA or aa), eliminating heterozygous genotypes. Therefore, the SNPs in these datasets are inherently binary-coded, without the heterozygous category. 
To align with the framework of our proposed method, we relabel $\{0,1\}$ to $\{-1,1\}$. \begin{revised}All tuning and performance reporting are conducted out-of-sample using 5-fold cross-validation.\end{revised} The goal of these analyses is to identify significant gene CMEs, which quantify the additive or dominant effect of one genetic variant conditional on the presence or absence of another genetic variant. 

\begin{table}[t]
\caption{Selection performance and predictive performance of six methods in the gene association study.}\label{tab:real_res}
\centering
\renewcommand{\arraystretch}{0.7}
\renewcommand{\cellset}{\renewcommand{\arraystretch}{0.5}\setlength{\tabcolsep}{3pt}}
\begin{tabular}{lrrrrrr}
    \toprule
     & \makecell[tl]{Adaptive\\ cmenet} & cmenet & Lasso & MCP & GEL & hierNet\\ 
    \toprule
    \textbf{(A) Maize flowering time} & & & & & & \\
    \# of selected effects & 19 & 24 & 43 & 12 & 21 & 66 \\
    \makecell[tl]{\# of significant selected \\effects ($p<0.05$)}  & 14 & 5 & 11 & 11 & 10 & 16 \\
    MSPE & 3.83 & 5.47 & 4.95 & 6.38 & 5.81 & 4.68 \\
    \midrule
    \textbf{(B) Arabidopsis thaliana} & & & & & & \\
    \# of selected effects & 9 & - & 13 & 11 & 5 & 93 \\
    \makecell[tl]{\# of significant selected \\effects ($p<0.05$)}  & 7 & - & 4 & 1 & 5 & 29 \\
    Misclassification Rate & 0.1 & - & 0.1 & 0.13 & 0.3 & 0.01 \\
    \bottomrule
\end{tabular}                    
\end{table}

In both studies, we compare the proposed approach to the cmenet, Lasso, MCP, GEL, and hierNet (R package \texttt{hierNet} V1.9; \citealp{hierNet}). The hierNet method \citep{bien2013lasso} is a widely used method for selecting two-way interactions (2FI) in high-dimensional data.  By comparing the adaptive cmenet to the hierNet, we aim to evaluate its practical advantage in identifying interaction relationships between genes. Note that the true underlying active effects are typically unknown in these real-world studies.
Here we focus on assessing whether the estimated model can select a smaller number of effects while maintaining high prediction accuracy \citep{stoica1982parsimony}. 
Such a criteria is crucial in genomic studies, where overly complex models with excessive effects may lead to reduced interpretability and overfitting. A model that selects fewer effects with comparable or better prediction accuracy is more likely to capture meaningful biological relationships while avoiding spurious findings. 
Additionally, we report the p-values of the selected effects using regular (generalized) linear regression to validate the relevance of the selected effects. This step can provide further evidence for the method’s ability to identify biologically meaningful associations. 

\begin{table}[t]
\caption{P-values of some selected effects for each method from a linear regression in Maize flowering time study.}
\centering
\renewcommand{\cellset}{\renewcommand{\arraystretch}{0.5}\setlength{\tabcolsep}{3pt}}
\resizebox{\textwidth}{!}{
\begin{tabular}{lrrrrrrrrr}
\toprule
\textbf{Method} & \textbf{$g1|g17-$} & \textbf{$g1|g39-$} & \textbf{$g9|g10-$} & \textbf{$g11|g38+$} & \textbf{$g12|g38-$} & \textbf{$g14|g38+$} & \textbf{$g15|g35+$} & \textbf{$g18|g33-$} & \textbf{$g27|g26-$} \\
\midrule
\makecell[tl]{Adaptive\\ cmenet} & - & 0.0004 & 0.0183 & 0.0000 & 0.0351 & 0.0042 & 0.0003 & 0.0062 & 0.0143 \\
\texttt{cmenet} & - & 0.1214 & 0.1281 & 0.0091 & 0.0124 & 0.0966 & 0.0007 & 0.1629 & 0.0180 \\
Lasso & 0.0392 & 0.0953 & 0.1951 & 0.0043 & 0.1050 & 0.0790 & 0.0372 & 0.0343 & 0.0200 \\
MCP & 0.2573 & 0.0544 & 0.0196 & 0.0333 & 0.0646 & 0.8667 & 0.0022 & 0.0133 & 0.0242 \\
GEL & 0.0085 & 0.0483 & 0.0049 & 0.0029 & 0.1048 & 0.0001 & 0.0001 & 0.0008 & 0.0081 \\
hierNet & 0.0259 & 0.0510 & 0.0002 & 0.0057 & 0.0000 & 0.0259 & 0.0002 & 0.0001 & 0.0001 \\
\bottomrule
\end{tabular}
}
\label{tab:maize_p}
\end{table}

The first study focuses on maize flowering time, a continuous variable measuring the number of days to flowering. We analyze a subset of the dataset from \cite{buckler2009genetic} containing $n=150$ observations, and $p=40$ main effects represented by binary SNP markers. Our proposed method demonstrates superior performance compared to several established methods. As shown in Table \ref{tab:real_res}(a), the adaptive cmenet identifies 14 significant effects out of the 19 selected, maintaining a lower MSPE than the hierNet. While the hierNet selects more significant effects, it also includes more insignificant ones, resulting in a higher MSPE. On the other hand, the MCP method selects fewer total effects (most of which are significant) but still exhibits a higher MSPE than the adaptive cmenet. These results demonstrate that the proposed method can efficiently capture key gene interactions while minimizing prediction error. Additionally, the p-value analysis of selected effects in Table \ref{tab:maize_p} reveals that adaptive cmenet identifies more significant effects than other methods, with considerable overlap between the significant effects selected by the adaptive cmenet and the hierNet.

\begin{table}[t]
\caption{The p-values of some selected effects for each method from a logistic regression in Arabidopsis thaliana study.}
\centering
\renewcommand{\cellset}{\renewcommand{\arraystretch}{0.5}\setlength{\tabcolsep}{3pt}}
\begin{tabular}{lrrrrrr}
\toprule
 & \makecell[tl]{Adaptive\\ cmenet}  & Lasso & MCP & GEL & hierNet & \makecell[tl]{Univariate\\ regression} \\
\midrule
$2:5804076|2:8750047+$ &  0.0069 & 0.0076 & - & 0.0000 & 0.0214 & 0.0007\\
$2:8750047|5:1931167+$ & 0.0337 & 0.0960 & - & 0.0004 & 0.0197 &0.0005 \\
$2:18752840|3:4818602+$ & 0.0122 & 0.0298 & - & 0.0008 & 0.0077 & 0.0066 \\
$2:18752840|4:11580143-$ & 0.0096 & 0.0133 & - & 0.0011 & 0.0276 & 0.0006 \\
$3:7392098|3:21632608+ $ & 0.0028 & 0.0491 & - & - &  0.3579 & 0.0034 \\
$4:17038400|2:9256095+ $ & 0.5980 & 0.0502 & - & - & 0.0020 &  0.0012 \\
$4:17038400|5:1931167-$ & 0.0314 & 0.2270 & - & - & 0.0197 & 0.0013\\
$5:20409250|2:9256095-$ & 0.0063 & 0.5506 & - & - & 0.0020 & 0.0153\\
\bottomrule
\end{tabular}
\label{tab:RA_p}
\end{table}

The second study investigates chlorosis phenotypes in Arabidopsis thaliana utilizing data from a genome-wide association study (GWAS) conducted by \citet{atwell2010genome}. Arabidopsis thaliana is a widely distributed, predominantly self-fertilizing model plant that exhibits substantial genetic variation for numerous adaptively significant traits \citep{atwell2010genome}. The dataset includes genotypic information for 250,000 single-nucleotide polymorphisms (SNPs) and phenotypic measurements for 107 traits across 199 strains, with data publicly available at \url{https://cynin.gmi.oeaw.ac.at/home/resources/atpolydb}. In terms of the visual chlorosis phenotype, assessed at 16°C after five weeks of growth, 84 strains exhibit chlorosis, while 92 do not. For this analysis, a subset of the dataset is selected, consisting of 100 strains (55 with visual chlorosis and 45 without), with $p=63$ main effects corresponding to binary SNP markers. The response variable is binary, representing the presence or absence of visual chlorosis. Consider Table \ref{tab:real_res}(B), the adaptive cmenet demonstrates superior performance compared to other methods by identifying a higher number of significant effects while maintaining a competitive misclassification rate. In contrast, the hierNet method selects the highest number of significant effects but also includes a larger number of insignificant effects, resulting in an overfitted model. The MCP method selects a comparable number of effects to adaptive cmenet; however, only one of them is statistically significant, and notably, it does not identify any effects that overlap with those selected by other methods. While the GEL method identifies fewer effects, of which five are significant, it exhibits higher misclassification rates. The adaptive cmenet achieves a better balance by selecting a greater number of significant dominant effects while minimizing the inclusion of insignificant effects, effectively combining selection accuracy with predictive reliability. The p-value analysis of selected effects in Table \ref{tab:RA_p} shows that both the adaptive cmenet method and the hierNet method consistently identify overlapping significant dominant effects. Notably, all CMEs identified by the adaptive cmenet are validated as statistically significant in the univariate regression model, further highlighting the superior selection performance of the proposed method.

\begin{revised} 
To investigate biological interpretations for the selected CMEs (Conditional Main Effects), we mapped each associated SNP to its nearest TAIR10 gene (within ±10 kb) and functionally annotated the genes using Gene Ontology, KEGG, and AraCyc. Intriguingly, the top CMEs frequently pair loci whose annotated functions suggest a conditional relationship relevant to chlorosis.
For instance, we detected a CME ($4:17038400|5:1931167-$) between APX5 (AT4G35970) and 2-Cys Peroxiredoxin B (AT5G06290). Both enzymes are critical components of ROS detoxification and redox buffering in photosynthetic tissues. The negative polarity of this CME indicated that the lack of 2-Cys Prx B removes its protective or regulatory function, making the APX5-associated genetic variation a more critical factor in determining the chlorotic phenotype. This finding is consistent with the compensation mechanism between the ascorbate peroxidase and peroxiredoxin redox pathways: loss of 2-Cys Prx activity can be compensated by other chloroplast antioxidant systems, most notably the ascorbate–glutathione cycle acting with superoxide dismutase, so that reliance shifts toward APX-linked routes \citep{puerto2013overoxidation}. This example elaborate the contribution of CMEs beyond traditional interactions: they capture context-dependent effects, where the influence of one locus emerges only in the presence (or absence) of a second, biologically related locus.
\end{revised} 

\section{Discussion}
\label{sec:conc}

This work has focused on adaptive bi-level variable selection of conditional mean effects for generalized linear models. The adaptive penalty structure used in the proposed method can enable a more flexible and adaptable selection of CMEs. The proposed adaptive cmenet enhances the ability to capture important conditional relationships while reducing the coupling constraints. 

Here we would like to remark that there are several aspects of improving the proposed method. 
First, similar to the IRLS algorithm, the algorithm for parameter estimation in Section 4 may not converge to a global optimal solution with respect to the actual objective function. 
It will be interesting to further investigate the properties of numerical convergence of the parameter estimation in the proposed method. 
Nevertheless, the proposed approach demonstrates robust performance in numerical studies.
Second, while the adaptive cmenet incorporates additional penalization to account for group structures such as cousins and siblings, the choice of initial weights could limit its ability to fully exploit these structures. Thus, alternative weighting schemes tailored to the CME framework may further improve the performance of adaptive cmenet.
\begin{revised}Third, while the similar theoretical framework for adaptive cmenet can be extended to canonical-link Poisson and Gamma regressions, the computational details are case-specific: variance functions (Poisson $\text{Var}=\mu$; Gamma $\text{Var}=\phi\mu^2$) and working weights alter curvature, step sizes, and numerical safeguarding. Cross-validation proceeds with the appropriate deviance loss, and warm starts/active-set updates remain effective, but require these case-specific investigations.\end{revised} 
Fourth, it is important to further improve the computational efficiency of the algorithm for scalability. 
Last, one can consider extending the proposed methodology to handle CMEs for multi-level or continuous variables, broadening its applicability to diverse datasets and experimental settings. 
Finally, given the origins of CMEs in experimental design, one can explore the use of CME in optimal experimental design and sequential design.

\section*{Supplementary Materials}
The Supplement.pdf contains additional methodology, proofs, algorithm details, and extended results.
The Simulation.zip contains R scripts used to reproduce simulations and figures reported in the manuscript.

\section*{Acknowledgments} 
The authors thank the Editor, the Associate Editor, and anonymous referees for their insightful comments and constructive suggestions that greatly improved the quality and clarity of this work.

\section*{Declaration of interest statement}
The authors report no conflict of interest.

\bibliographystyle{apalike}
\bibliography{bibliography.bib}

\newpage
\setcounter{footnote}{0}

	\begin{center}
		{\Large \bf Supplementary Materials for \\
        ``Adaptive Bi-Level Variable Selection of Conditional Main Effects for Generalized Linear Models"}\\
		\vskip 10pt
	
	\end{center}

This supplementary material contains 
(i) Technical proofs for Proposition 1, Corollary 1.1, Lemma 1, and Theorem 1, (ii) the algorithm for adaptive cmenet in Section 4.2, and (iii) the sensitivity analysis for across effect size $\beta_{\text{CME}}$ in Example 2 (Section 5.2).

\setcounter{page}{1}
\setcounter{section}{1}
\setcounter{equation}{0}
\renewcommand{\theequation}{A.\arabic{equation}}
\appendix
\setcounter{figure}{0}
\renewcommand{\thefigure}{A.\arabic{figure}}    

\subsection*{Proof of Proposition 1}
$Q(\boldsymbol{\beta})$ is non-differentiable at $\boldsymbol{\beta}=\mathbf{0_p}$, but it is semi-differentiable at first and second order. That is, it possesses a directional derivative for all points $\boldsymbol{\beta}$ and in all directions $\mathbf{u} \in \mathbb{R}^p$ ($\Vert \mathbf{u}\Vert=1$) for $\boldsymbol{\beta}$. We follow the similar approach as Proposition 1 of \cite{mak2019cmenet}. Let $\omega_j$ be the individual weights for all coefficients $\beta_j$, $j=1,\dots,p'$ and define $\omega_{\max}=\max_{1\leq j \leq p'}\omega_j$. The second derivative of $Q(\boldsymbol{\beta})$ in direction $\mathbf{u}$ is:
\begin{equation}
    \nabla^2_{\mathbf{u}}Q(\boldsymbol{\beta})=\frac{1}{n}\mathbf{u}^T{\mathbf{X}}^{T}\mathbf{W}\mathbf{X}\mathbf{u}+\nabla^2_{\mathbf{u}}\tilde{P}_S(\boldsymbol{\beta})+\nabla^2_{\mathbf{u}}\tilde{P}_C(\boldsymbol{\beta}),
\label{prop1:eq2}
\end{equation}
where $\nabla^2_{\mathbf{u}}\tilde{P}_\mathcal{G}(\boldsymbol{\beta})=\mathbf{u}^T \nabla^2_{\boldsymbol{\beta}}\tilde{P}_\mathcal{G}(\boldsymbol{\beta})\mathbf{u}$ for $\mathcal{G} \in \{\mathcal{S}, \mathcal{C}\}$. Without loss of generalization, the elements of the Hessian matrix $H_{\boldsymbol{\beta}}\overset{\underset{\mathrm{def}}{}}{=}\nabla^2_{\boldsymbol{\beta}}\tilde{P}_\mathcal{S}(\boldsymbol{\beta})$ are given by:
\begin{equation}
\resizebox{0.9\hsize}{!}{ 
$\begin{aligned}
&\frac{\partial ^2}{\partial \beta_{j|m+}\partial \beta_{h|n+}}\tilde{P}_\mathcal{S}(\boldsymbol{\beta}) = \\
&\left\{
    \begin{aligned}
        &{f^{(j)}_{o,\mathcal{S}}}''\Biggl\{ \sum_{k\in \mathcal{S}(j)} \omega_k f^{(j)}_{i,\mathcal{S}}(\beta_k) \Biggr\}({f^{(j)}_{i,\mathcal{S}}}'(\beta_{j|m+}))^2\omega_{j|m+}^2+{f^{(j)}_{o,\mathcal{S}}}'\Biggl\{ \sum_{k\in \mathcal{S}(j)} \omega_k f^{(j)}_{i,\mathcal{S}}(\beta_k) \Biggr\}{f^{(j)}_{i,\mathcal{S}}}''(\beta_{j|m+})\omega_{j|m+},  &&\text{if } j=h, m=n,\\
        &{f^{(j)}_{o,\mathcal{S}}}''\Biggl\{ \sum_{k\in \mathcal{S}(j)} \omega_k f^{(j)}_{i,\mathcal{S}}(\beta_k) \Biggr\}{f^{(j)}_{i,\mathcal{S}}}'(\beta_{j|m+}){f^{(j)}_{i,\mathcal{S}}}'(\beta_{j|n+})\omega_{j|m+}\omega_{j|n+}, &&\text{if } j=h, m \neq n,\\
        &0, &&\text{if } j \neq h.
    \end{aligned}
     \right.
\end{aligned}$
}
\label{prop1:eq3}
\end{equation}
Hence, it is easy to show that $$\nabla^2_{\mathbf{u}}\tilde{P}_\mathcal{G}(\boldsymbol{\beta} )=\sum_{i=1}^{p'}\sum_{j=1}^{p'}\{H_{\boldsymbol{\beta}}\}_{i,j}u_iu_j \geq -\tau\exp{(0)}\omega^2_{\max}+\lambda_{\mathcal{G}(j)}\exp{(0)}(-\frac{1}{\lambda_{\mathcal{G}(j)}\gamma})\omega_{\max}=-\tau \omega^2_{\max}-\frac{\omega_{\max}}{\gamma},$$
for all $\mathbf{u}$ and $\boldsymbol{\beta}$. Moreover, let $\mathbf{u} = \sum_{i=1}^{p'} u_i \mathbf{e}_i$, where \(\mathbf{e}_i\) represents the unit eigenvectors of \(\mathbf{X}^T \mathbf{W} \mathbf{X}\), satisfying the eigenvalue equation  $\mathbf{X}^T \mathbf{W} \mathbf{X} \mathbf{e}_i = \zeta_i \mathbf{e}_i$. Then, we have  
 $\mathbf{u}^T (\mathbf{X}^T \mathbf{W} \mathbf{X}) \mathbf{u} = \sum_{i=1}^{p'} u_i^2 \zeta_i \geq \zeta_{\min}$. Therefore,
$$\nabla^2_{\mathbf{u}}Q(\boldsymbol{\beta}) \geq \frac{\zeta_{\min}(\mathbf{X}^T \mathbf{WX})}{n}-2(\tau \omega^2_{\max}+\frac{\omega_{\max}}{\gamma}),$$ which is strictly positive when $\tau+\frac{1}{\gamma \omega_{\max}} < \frac{\zeta_{\min}(\mathbf{X^T WX})}{2n \omega^2_{\max}}$.

\subsection*{Proof of Corollary 1.1}
From Proposition 1, the objective function \( Q(\boldsymbol{\beta}) \) is strictly convex if the following condition holds $\tau+\frac{1}{\gamma \omega_{\max}} < \frac{\zeta_{\min}(\mathbf{X}^T \mathbf{W} \mathbf{X})}{2n \omega^2_{\max}} $, where \( \mathbf{W} = \text{diag}(w_1, w_2, \dots, w_n) \) is a diagonal matrix of weights. In the case of logistic regression, the weight terms are defined as
$w_i = \pi_i(1 - \pi_i)$, where $\pi_i = \frac{\exp(\mathbf{x}_i^T \boldsymbol{\beta})}{1+\exp(\mathbf{x}_i^T \boldsymbol{\beta})}$. It follows that \( w_i \leq 1/4 \), with equality attained when \( \pi_i = 1/2 \). 

Furthermore, utilizing the property \( \Vert \mathbf{x}_j\Vert_2^2 = n \), we establish that the coordinate-wise objective function \( Q_j(\boldsymbol{\beta}_j) \) remains strictly convex for all \( j = 1, \dots, p' \) whenever the condition $\tau+\frac{1}{\gamma \omega_{\max}} < \frac{1}{8 \omega^2_{\max}}$ is satisfied.

\subsection*{Proof of Lemma 1}
By first-order Taylor expansion, since $\eta_{\lambda,\tau}(\theta)$ is a concave and differentiable function, it is easy to show:
\begin{equation*}
    \eta_{\lambda_{\mathcal{G}},\tau}(\boldsymbol{\beta}_{\mathcal{G}}|\boldsymbol{\beta}_{\mathcal{G}}^*) \leq \eta_{\lambda_{\mathcal{G}},\gamma}\{\Vert \boldsymbol{\beta}^*_{\mathcal{G}} \Vert_{\lambda_{\mathcal{G}},\omega,\gamma}\}+\Delta^*_{\mathcal{G}}\times\{\Vert \boldsymbol{\beta}_{\mathcal{G}} \Vert_{\lambda_{\mathcal{G}},\omega,\gamma}-\Vert \boldsymbol{\beta}^*_{\mathcal{G}} \Vert_{\lambda_{\mathcal{G}},\omega,\gamma}\},
\end{equation*}
where the inequality holds due to the concavity of $\eta$ on $[0,\infty)$. Therefore, by direct substitution, $\tilde{Q}_{j|k+}(\beta_{j|k+}|\boldsymbol{\beta}^*)$ in \eqref{equ8} majorizes $Q_{j|k}(\beta_{j|k+})$.

\subsection*{Proof of Theorem 1}
To facilitate the analysis, for any effect coefficient $\beta_j$, $j=1,\dots,p'$, let $\lambda_1$ and $\lambda_2$ denote the tuning parameters associated with the sibling group and cousin group, respectively. The corresponding penalty slopes are denoted by $\Delta_1$ and $\Delta_2$. Consider optimization problem:
\begin{equation}
    \hat{\beta_j}=\min_{\beta_j}\Biggl[\frac{1}{2n}(\tilde{\mathbf{y}}-\mathbf{X}\boldsymbol{\beta})^{T}\mathbf{W}(\tilde{\mathbf{y}}-\mathbf{X}\boldsymbol{\beta})+ \Delta_1 \omega_{j} g_{\lambda_1,\gamma}(\beta_j)+ \Delta_2 \omega_{j} g_{\lambda_2,\gamma}(\beta_j)\Biggr]
\label{theorm1:eq1}
\end{equation}
This equivalent to 
\begin{equation}
 \begin{cases}
\min_{\beta_j}L(\beta_j)=\frac{1}{2n}(\tilde{\mathbf{y}}-\mathbf{X}\boldsymbol{\beta})^{T}\mathbf{W}(\tilde{\mathbf{y}}-\mathbf{X}\boldsymbol{\beta}) \\
s.t., \omega_{j}g_{\lambda_1,\gamma}(\beta_j)\leq 0, \omega_{j}g_{\lambda_2,\gamma}(\beta_j)\leq 0
\end{cases} 
\label{theorm1:eq2}
\end{equation}
Taking the derivative of the objective function with respect to $\beta_j$, we obtain:
\begin{equation*} 
\begin{aligned}
\frac{\partial L(\beta_j)}{\partial \beta_j} &= \frac{1}{2n} \left(-2 \mathbf{x}_j^T \mathbf{W} \tilde{\mathbf{y}} + 2 \mathbf{x}_j^T \mathbf{W} \mathbf{X} \boldsymbol{\beta} \right) \\
&=-\frac{1}{n} \mathbf{x}_j^T \mathbf{W}(\tilde{\mathbf{y}}-\mathbf{X}_{-j}\boldsymbol{\beta}_{-j}) + \frac{1}{n} \mathbf{x}_j^T \mathbf{W} \mathbf{x}_j \beta_j\\
&= -z_j + v_j \beta_j,
\end{aligned}
\end{equation*}
where $z_j = \frac{1}{n} \mathbf{x}_j^T \mathbf{W}(\tilde{\mathbf{y}}-\mathbf{X}_{-j}\boldsymbol{\beta}_{-j})$, and $v_j = \frac{1}{n} \mathbf{x}_j^T \mathbf{W} \mathbf{x}_j$.
For the constraint function $ g_{\lambda,\gamma}(\beta_j)$, the derivative with respect to $\beta_j$ is:
\begin{equation*} 
\partial g_{\lambda,\gamma}(\beta_j) = \begin{cases} 0, & \text{if } |\beta_j| > \lambda\gamma, \\
\text{sgn}(\beta_j) \left(1 - \frac{|\beta_j|}{\lambda\gamma} \right), & \text{if } |\beta_j| \leq \lambda\gamma, \\ 
[ -1, 1 ], & \text{if } \beta_j = 0. \end{cases} 
\end{equation*}
Applying the Karush-Kuhn-Tucker (KKT) conditions, we derive the optimality conditions:
\begin{itemize}
    \item[1.] Stationarity condition: 
    \begin{equation} 
    0 \in -z_j + v_j \beta_j +\Delta_1 \omega_{j} \partial g_{\lambda_1,\gamma}(\beta_j)+ \Delta_2 \omega_{j} \partial g_{\lambda_2,\gamma}(\beta_j).
    \label{kkt1}
    \end{equation}
    \item[2.] Complementary slackness condition: $ \Delta_1 \omega_{j} g_{\lambda_1,\gamma}(\beta_j) = 0, \ \Delta_2 \omega_{j} g_{\lambda_2,\gamma}(\beta_j) = 0 $.
    \item[3.] Primal feasibility condition: $ \omega_{j}g_{\lambda_1,\gamma}(\beta_j)\leq 0, \  \omega_{j}g_{\lambda_2,\gamma}(\beta_j)\leq 0$.
    \item[4.] Dual feasibility condition: $ \Delta_1 \geq 0, \Delta_2 \geq 0 $.
\end{itemize}
Without loss of generality, assume $z_j \geq 0$. Let $\lambda_{(1)}=\max(\lambda_{1},\lambda_{2})$ and $\lambda_{(2)}=\min(\lambda_{1},\lambda_{2})$, with $\Delta_{(1)}$ and $\Delta_{(2)}$ its corresponding slopes. Consider the same four cases for $z_j$ as presented in \eqref{equ9}:
\begin{itemize}
 \item[1.] If $\hat{\beta}_j \geq \lambda_{(1)}\gamma$, then from the stationarity condition \eqref{kkt1}, $0 \in -z_j + v_j \hat{\beta}_j$, which implies: $\hat{\beta}_j = \frac{z_j}{v_j}$. Thus, $z_j \geq v_j\lambda_{(1)}\gamma$. Since \eqref{theorm1:eq1} is strictly convex, $\hat{\beta}_j = \frac{z_j}{v_j}$ is the unique solution to \eqref{theorm1:eq1}.
 \item[2.] If $\lambda_{(2)}\gamma \leq \hat{\beta}_j < \lambda_{(1)}\gamma$, then from the stationarity condition \eqref{kkt1}, $0 \in -z_j + v_j \hat{\beta}_j + \Delta_{(1)}\omega_j(1-\frac{\hat{\beta}_j}{\lambda_{(1)}\gamma})$, which implies: $\hat{\beta}_j = \frac{z_j-\Delta_{(1)}\omega_j}{v_j-\frac{\Delta_{(1)}\omega_j}{\lambda_{(1)}\gamma}}$. Thus, $v_j\lambda_{(2)}\gamma + \Delta_{(1)}\omega_j(1-\frac{\lambda_{(2)}}{\lambda_{(1)}})\leq z_j < v_j\lambda_{(1)}\gamma$. Since \eqref{theorm1:eq1} is strictly convex, $\hat{\beta}_j$ is the unique solution.
 \item[3.] If $0 \leq \hat{\beta}_j < \lambda_{(2)}\gamma$, then from the stationarity condition \eqref{kkt1}, $0 \in -z_j + v_j \hat{\beta}_j + \Delta_{(1)}\omega_j(1-\frac{\hat{\beta}_j}{\lambda_{(1)}\gamma}) + \Delta_{(2)}\omega_j(1-\frac{\hat{\beta}_j}{\lambda_{(2)}\gamma})$, which implies: $\hat{\beta}_j = \frac{z_j-\Delta_{(1)}\omega_j-\Delta_{(2)}\omega_j}{v_j-\frac{\Delta_{(1)}\omega_j}{\lambda_{(1)}\gamma}-\frac{\Delta_{(2)}\omega_j}{\lambda_{(2)}\gamma}}$. Thus, $\Delta_{(1)}\omega_j + \Delta_{(2)}\omega_j \leq z_j < v_j\lambda_{(2)}\gamma + \Delta_{(1)}\omega_j(1-\frac{\lambda_{(2)}}{\lambda_{(1)}})$. Since \eqref{theorm1:eq1} is strictly convex, $\hat{\beta}_j$ is the unique solution to \eqref{theorm1:eq1}.
 \item[4.] If $\hat{\beta}_j = 0$, then from the stationarity condition \eqref{kkt1}, $0 \in -z_j + \Delta_{(1)}\omega_j[-1,1] + \Delta_{(2)}\omega_j[-1,1]$, which implies: $0 \leq z_j < \Delta_{(1)}\omega_j + \Delta_{(2)}\omega_j$. Since \eqref{theorm1:eq1} is strictly convex, $\hat{\beta}_j = 0$ is the unique solution to \eqref{theorm1:eq1}.
\end{itemize}

\subsection*{Proof of Proposition 2}
Proposition 2 can be proved in a similar way as Proposition 3 of \cite{breheny2015group}.

\subsection*{Algorithm for the adaptive cmenet}

\begin{algorithm}[H]
\caption{Adaptive cmenet: An algorithm for adaptive bi-level cmenet selection}\label{alg:glmcmenet}
\begin{algorithmic}[1]
\Function{adaptive cmenet}{$\mathbf{X},\mathbf{y},\lambda_s,\lambda_c,\gamma,\tau,\boldsymbol{\beta}=\mathbf{0_{p^{'}}},\boldsymbol{\hat{\beta}}^{ridge}$} 
\State{Initialize} $\beta_0 = \log(\bar{y}/(1-\bar{y})),\eta_i=\log(\bar{y}/(1-\bar{y}))$, $\Omega_{S(j)}=(\|\boldsymbol{\hat{\beta}}^{ridge}_{S(j)}\|_1 + \frac{1}{n})^{-1}$, $\Omega_{C(j)}=(\|\boldsymbol{\hat{\beta}}^{ridge}_{C(j)}\|_1 + \frac{1}{n})^{-1}$, $\Delta_{S(j)}=\lambda_s$, $\Delta_{C(j)}=\lambda_c$, $\lambda_{S(j)}=\lambda_s\Omega_{S(j)}$, $\lambda_{C(j)}=\lambda_s\Omega_{C(j)}$, for $j=1,2,...,p$; $\boldsymbol{\omega}=\{(|{\hat{\beta}}^{ridge}_j| + \frac{1}{n})^{-1}\}_{j=1}^{p'}$
\Repeat
 \State Calculate $\pi_i \gets \frac{\exp(\eta_i)}{1+\exp(\eta_i)}$, $w_i \gets \pi_i(1-\pi_i)$, $r_i \gets \frac{y_i-\pi_i}{w_i}$, for $i=1,2,...,n$
 \State Update $\beta^{(0)}_0 \gets \beta_0$, $\beta_0 \gets \frac{\mathbf{1_n^{T}Wr}}{\mathbf{1_n^{T}W1_n}}+\beta^{(0)}_0$ \Comment{For intercept}
 \State $\mathbf{r} \gets \mathbf{r}-\mathbf{1}_n(\beta_0-\beta^{(0)}_0)$, $\boldsymbol{\eta} \gets \boldsymbol{\eta}+\mathbf{1}_n(\beta_0-\beta^{(0)}_0)$
  \For{$j=1,\dots,p$} \Comment{For all MEs}
    \State Calculate $v_j \gets \frac{1}{n}\mathbf{x}_j^T\mathbf{Wx}_j$, $z_j \gets \frac{1}{n}\mathbf{x}_j^T\mathbf{Wr}+v_j\beta^{(0)}_j$
    \State Update $\beta^{(0)}_j \gets \beta_j$, $\beta_j \gets S_{\lambda_{S(j)},\lambda_{C(j)}}(z_j,\Delta_{S(j)},\Delta_{C(j)})$
    \State $\mathbf{r} \gets \mathbf{r}-\mathbf{x}_j(\beta_j-\beta^{(0)}_j)$, $\boldsymbol{\eta} \gets \boldsymbol{\eta}+\mathbf{x}_j(\beta_j-\beta^{(0)}_j)$
    \State $\Delta_{S(j)} \gets \Delta_{S(j)}\exp\{-\frac{\tau}{\lambda_{S(j)}}w_j[g_{\lambda_{S(j)},\gamma}(\beta_j)-g_{\lambda_{S(j)},\gamma}(\beta^{(0)}_j)]\}$
    \State $\Delta_{C(j)} \gets \Delta_{C(j)}\exp\{-\frac{\tau}{\lambda_{C(j)}}w_j[g_{\lambda_{C(j)},\gamma}(\beta_j)-g_{\lambda_{C(j)},\gamma}(\beta^{(0)}_j)]\}$
  \EndFor
 \For{$j=1,\dots,p$ and $k=1,\dots,p$ do} \Comment{For all CMEs}
    \State Calculate $v_{j|k+} \gets \frac{1}{n}\mathbf{x}_{j|k+}^T\mathbf{Wx}_{j|k+}$, $z_j \gets \frac{1}{n}\mathbf{x}_{j|k+}^T\mathbf{Wr}+v_{j|k+}\beta^{(0)}_{j|k+}$
    \State Update $\beta^{(0)}_{j|k+} \gets \beta_{j|k+}$, $\beta_{j|k+} \gets S_{\lambda_{S(j)},\lambda_{C(k)}}(z_{j|k+},\Delta_{S(j)},\Delta_{C(k)})$
    \State $\mathbf{r} \gets \mathbf{r}-\mathbf{x}_{j|k+}(\beta_{j|k+}-\beta^{(0)}_{j|k+})$, $\boldsymbol{\eta} \gets \boldsymbol{\eta}+\mathbf{x}_{j|k+}(\beta_{j|k+}-\beta^{(0)}_{j|k+})$
    \State $\Delta_{S(j)} \gets \Delta_{S(j)}\exp\{-\frac{\tau}{\lambda_{S(j)}}w_{j|k+}[g_{\lambda_{S(j)},\gamma}(\beta_{j|k+})-g_{\lambda_{S(j)},\gamma}(\beta^{(0)}_{j|k+})]\}$
    \State $\Delta_{C(k)} \gets \Delta_{C(k)}\exp\{-\frac{\tau}{\lambda_{C(k)}}w_{j|k+}[g_{\lambda_{C(k)},\gamma}(\beta_{j|k+})-g_{\lambda_{C(k)},\gamma}(\beta^{(0)}_{j|k+})]\}$
  \EndFor
\Until{$\mathbf{\beta}$ converges}
\State \Return the converged coefficient vector $\boldsymbol{\beta}$
\EndFunction
\end{algorithmic}
\end{algorithm}

\subsection*{Effect of signal-to-noise ratio}
\begin{sidewaysfigure}
    \centering
     \includegraphics[width=1\textwidth]{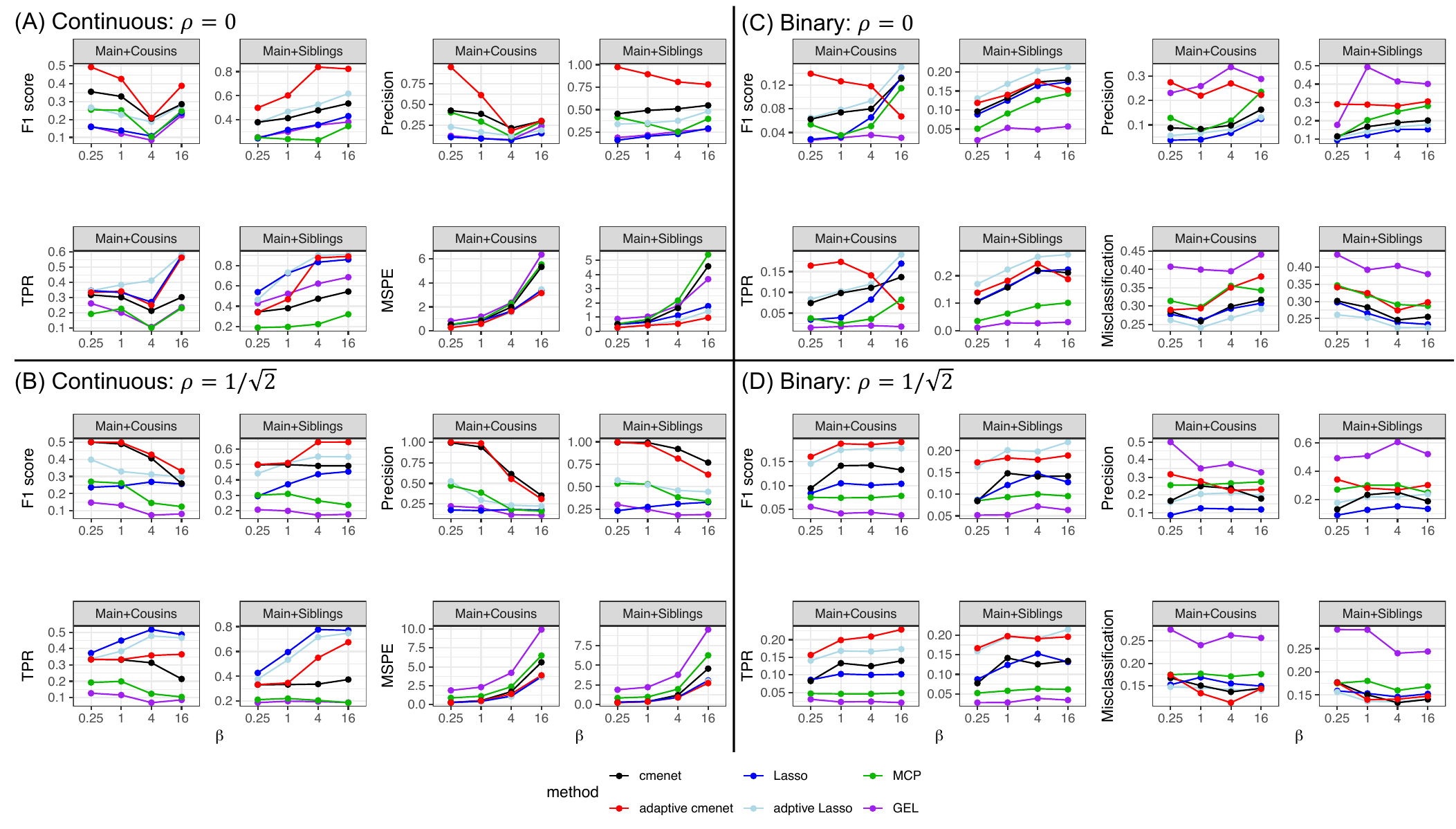}   
    \caption{\begin{revised}Sensitivity analysis across effect size $\beta_{\text{CME}}$ (Example 2; $n=50$, $p=20$; fixed $\beta_{\text{ME}}=5$): 4 active groups with 2 active effects in each group. Panels (A)–(D) show: (A) continuous, uncorrelated ($\rho=0$); (B) continuous, correlated ($\rho=1/\sqrt{2}$); (C) binary, uncorrelated ($\rho=0$); and (D) binary, correlated ($\rho=1/\sqrt{2}$).
    The $x$-axis varies $\beta_{\mathrm{CME}}$; the $y$-axis reports $F_{1}$, Precision, True Positive Rate (TPR), and MSPE (continuous) or Misclassification Rate (binary).
    Columns correspond to main+cousins (left) and main+siblings (right).\end{revised}}
    \label{fig:exp2_beta_sensitivity}
\end{sidewaysfigure}

\begin{revised}
Figure \ref{fig:exp2_beta_sensitivity} summarizes how selection and prediction change as the CME effect size $\beta_{\text{CME}}$ increases (with fixed $\beta_{\text{ME}}=5$) across continuous and binary outcomes, uncorrelated and correlated settings, and main+cousins versus main+siblings structures, following Scenario 2 in Section \ref{sim:second}. In the continuous panels, as $\beta_{\text{CME}}$ increases (i.e., the ratio $\beta_{\text{ME}}/\beta_{\text{CME}}$ decreases), F1 and TPR decline for the main+cousins design but increase for main+siblings; in both cases adaptive cmenet maintains the highest 
F1 overall. Precision remains the highest under $\rho=0$ and is comparable to non-adaptive cmenet under $\rho=1/\sqrt2$. MSPE rises with $\beta_{\text{CME}}$ because the response scale grows with signal strength; notably, adaptive cmenet shows the slowest MSPE growth among methods. In the binary panels, the same F1 or TPR patterns hold as $\beta_{\text{CME}}$ grows. Precision is lower than GEL but higher than other baselines in both correlation settings across all effect sizes. The misclassification rate remains lower or comparable to other methods across all settings, indicating consistently strong out-of-sample prediction performance. Overall, a larger $\beta_{\text{ME}}/\beta_{\text{CME}}$ (smaller relative $\beta_{\text{CME}}$) widens performance gaps, with our method showing the greatest advantage; when $\beta_{\text{ME}}/\beta_{\text{CME}}$ is smaller, the gaps narrow, yet our method still outperforms the alternatives. Note that in the continuous \& uncorrelated case, when MEs and CMEs are similar in magnitude (e.g., $\beta_{\text{ME}}=5$, $\beta_{\text{CME}}=4$), all methods perform worst among the examined ratios. When the MEs and CMEs have about the same strength, the model has trouble deciding which terms to keep. The signal is split between MEs and CMEs, so each looks only moderately strong on its own. As a result, many effects fall near the selection cutoff, making it easier to miss true signals or include wrong ones. Because there’s no correlation structure to help “link” related terms, the model cannot borrow information across features to stabilize selection. This leads to lower F1 or TPR and higher error measures, even though the overall signal size is not weak.
\end{revised}

\end{document}